\documentclass[fleqn,usenatbib]{mnras}

\usepackage{newtxtext,newtxmath}

\usepackage[T1]{fontenc}

\DeclareRobustCommand{\VAN}[3]{#2}
\let\VANthebibliography\thebibliography
\def\thebibliography{\DeclareRobustCommand{\VAN}[3]{##3}\VANthebibliography}

\usepackage{graphicx}	
\usepackage{amsmath}	

\title[Astrometry via Close Approach Events]{Astrometry via Close Approach Events: Applications to Main-Belt Asteroid (702) Alauda}

\author[Guo et al.]{
	B. F. Guo$^{1,3}$,
	Q. Y. Peng$^{1,3}$\thanks{E-mail: tpengqy@jnu.edu.cn},
	A. Vienne$^{2,3}$,
	X. Q. Fang$^{1,3}$
\\
$^{1}$Department of Computer Science, Jinan University, Guangzhou 510632, China\\
$^{2}$IMCCE, Observatoire de Paris, PSL Research University, CNRS, Sorbonne Universit\'e, Universit\'e de Lille, Paris 75014, France\\
$^{3}$Sino-French Joint Laboratory for Astrometry, Dynamics and Space Science, Jinan University, Guangzhou 510632, China}

\date{Accepted 2023 August 08; 2023 August 06; in original form 2023 July 12}

\pubyear{2023}

\begin{document}
\label{firstpage}
\pagerange{\pageref{firstpage}--\pageref{lastpage}}
\maketitle

\begin{abstract}
The release of \textit{Gaia} catalog is revolutionary to the astronomy of solar system objects. After some effects such as atmospheric refraction and CCD geometric distortion have been taken into account, the astrometric precision for ground-based telescopes can reach the level of tens of milli-arcseconds. If an object approaches a reference star in a small relative angular distance (less than 100 arcseconds), which is called close approach event in this work, the relative positional precision between the object and reference star will be further improved since the systematic effects of atmospheric turbulence and local telescope optics can be reduced. To obtain the precise position of a main-belt asteroid in an close approach event, a second-order angular velocity model with time is supposed in the sky plane. By fitting the relationship between the relative angular distance and observed time, we can derive the time of maximum approximation and calculate the corresponding position of the asteroid. In practice, 5 nights' CCD observations including 15 close approach events of main-belt asteroid (702) Alauda are taken for testing by the 1m telescope at Yunnan Observatory, China. Compared with conventional solutions, our results show that the positional precision significantly improves, which reaches better than 4 milli-arcseconds, and 1 milli-arcsecond in the best case when referenced for JPL ephemeris in both right ascension and declination.
\end{abstract}


\section{Introduction}

\par The astrometry of asteroids is helpful to reveal the formation and evolution of the solar system. For example, the precise positions of asteroids are essential to estimate their accurate orbits \citep{Milani2010book, Desmars2013AA}. The study of Yarkovsky effect also needs accurate positions of the asteroids \citep{Vigna2018AA, Greenberg2020AJ}. In addition, the astrometry of asteroids are meaningful to the defense of impact. \citet{Spoto2017AA} presented that astrometric accuracy affected the impact monitoring of near-Earth asteroids (NEAs), especially for short-arc orbital solutions. \citet{Siltala2020AA} found that the high-accuracy astrometry of an asteroid could reduce the uncertainty of mass estimate.

\par At present, most observations of asteroids are obtained via ground-based telescopes. The event of stellar occultation is one of the approach to obtain not only the size and shape of an asteroid, but also the accurate astrometry \citep{Ferreira2022AA}. For a main-belt asteroid (MBA), a well-observed stellar occultation under the right circumstances can establish the position of the shadow better than 1 km, which is equivalent to less than 0.5 milli-arcseconds \citep{Herald2020MNRAS}. However, if a stellar occultation by a given asteroid cannot be observed, one may obtain accurate astrometry from close approach events with Gaia stars, as described in this work. \textit{Gaia} catalog \citep{GaiaDR3} provides the accurate and precise positions of reference stars, which is revolutionary to the astrometry of solar system objects. Nonetheless, it is not easy to obtain the accurate and precise positions of asteroids. For fast-moving asteroid, long trails exist even if a short exposure time is adopted, which causes the loss of precision. \citet{Shao2014ApJ} developed the technique of synthetic tracking to obtain accurate positions for NEAs. Later, \citet{Zhai2018AJ} reported the precision can reach the level of 10 milli-arcseconds using this technique. Moreover, the CCD distortion influences the astrometry of asteroids \citep{Peng2012AJ, Wang2015MN}, which should be also taken account. In addition, the effect of differential color refraction (DCR) results in systematic positional errors towards zenith, which should be considered to obtain the accurate and precise positions \citep{Anderson2006AA, Velasco2016MNRAS}.

\par If we consider the effects of atmospheric turbulence and local telescope optics, the precision of astrometry will be further improved. An asteroid approaches a reference star in a small angular distance (usually less than 100 arcseconds), which is called an approach event in this work, during which these effects can be reduced. Similar to close approach event, appulse is also an event with the minimum apparent angular distance between two objects. One of the difference is that the two objects have even smaller angular distance during an appulse so that the extended part of the body (such as the atmosphere) is occulted as shown in \citet{Kammer2020AJ}. Another difference is that observations with less spans of angular distances are required (e.g. within the range of only several arcseconds) in an appulse like stellar occultations. The reasons for positional precision improvement in a close approach event mainly come from the local similar atmospheric turbulence and telescope optical effect. \citet{Morgado2016MN} used this method to determine the positional difference between two concerned planetary satellites. This method was successfully applied to the observations of Galilean satellites of Jupiter \citep{Morgado2019MN} and satellites of Uranus \citep{Santos2019MNRAS}. For the practice of Galilean satellites, the average internal precision can reach 11.3 milli-arcseconds within the angular distance of 30 arcseconds \citep{Morgado2019MN}, which is almost comparable to the positional precision from mutual phenomenon. \citet{Lin2019MN} found that precision premium occurred within the angular distance of 100 arcseconds and the improvement of the precision could be expressed as a sigmoidal function with the angular distance.

\par In this paper, we want to obtain higher positional precision of an asteroid after considering the effects of image quality, geometric distortion, atmospheric refraction (including DCR effect), atmospheric turbulence and local telescope optics. Meanwhile, we want to extend the study of the close approach events from two satellites \citep{Morgado2016MN, Morgado2019MN} to an asteroid with respect to its nearby reference star. In fact, the close approach event happens on all the solar system objects. In this work, considering the aperture size of the telescope used and the available observation time, we choose the typical objects MBAs for testing because of the high signal-noise ratios (SNRs), large angular velocity but less trailing. The main-belt asteroid (702) Alauda is appropriate for testing due to the enough calibrated stars in the field of view and several close approach events during one night (also see Section \ref{section:observations}).

\par The remainder of this paper is organized as follows. Section \ref{section:observations} will introduce the telescope, CCDs and observations. Section \ref{section:Method} will elaborate the reduction method for a main-belt asteroid in a close approach event. Section \ref{section: Results} will show the astrometric reduction results and analysis of main-belt asteroid (702) Alauda. Finally, the drawn conclusions will be presented in the last section.

\begin{table}
	\centering
	\small
	\caption{The specifications of CCDs used of the 1m telescope.}
	\begin{tabular}{ccc}
		\hline
		\hline
		\makebox[0.13\textwidth][c]{Parameter} &\makebox[0.11\textwidth][c]{CCD\#1}& \makebox[0.11\textwidth][c]{CCD\#2}  \\
		\hline
		Size of CCD array            & 2048 $\times$ 2048      & 4096 $\times$ 4112                  \\
		Size of pixel                & 13.5$\mu $ $\times $ 13.5$\mu $     &15$\mu $ $\times$ 15$\mu$   \\
		Approximate scale factor     &
		0\farcs209/pixel       & 0\farcs233/pixel              \\
		Field of view                & $7\farcm1 \times7\farcm1$                    & $16\farcm0 \times 16\farcm0$             \\
		Nights observed            & 2                                   & 3                                  \\
		\hline
	\end{tabular}
	\label{table:Instrumental Details}
\end{table}

\section{Observations}
\label{section:observations}
\par Five nights' CCD observations including 15 close approach events of asteroid (702) Alauda are taken by the 1m telescope at Yunnan Observatory, China (IAU code: 286). Totally, 1092 CCD frames are taken from Nov 11, 2020 to Feb 10, 2021. We exclude some observations ($\sim$2\% of the total) when they are taken with the nearby service lights turned on by mistake. For the telescope used, the diameter of the primary mirror is 101.6 cm, and the $ f $ ratio is 13.3. Two different CCDs are used during our observation. The specifications of the CCDs are listed in Table \ref{table:Instrumental Details}. As we aimed to minimize the effect of differential color refraction, according to \cite{Stone2002PASP} and \cite{Lin2020MNRAS}, we adopt the scheme that all the observations are taken with \textit {Cousins-I} filter and the observational zenith distances are less than ${60^\circ}$. To ensure the measured positional precision of Alauda, we empirically make its trailing less than 1/3 of seeing. The visual magnitude of Alauda is 12 $\sim$ 13 during the observation. The magnitude differences between Alauda and nearby reference stars (the visual magnitude of Alauda and \textit{Gaia-G} magnitude of stars) are less than 5. More details of the observations are shown in Table \ref{table:Observations Details}.

\section{Method}
\label{section:Method}

\subsection{Data Reduction}
\label{section:Data Reduction}
\par Firstly, all CCD frames are pre-processed by bias and flat-field corrections. Then, for each frame, we detect all the sources and measure their pixel coordinates with two-dimensional Gaussian centering algorithm. Specifically, for each image, one can refer to DAOPHOT program \citep{Stetson1987PASP} for detecting each source and deriving its initial position in pixel coordinate. To further measure the pixel coordinate of each source, we use the two-dimensional Gaussian centering algorithm. For more details of the centering algorithm, one can refer to eq.(3) in \citet{Lin2021ApSS}. Secondly, we match these sources with the stars in \textit{Gaia} DR3 catalog \citep{GaiaDR3} and calculate their topocentric astrometric positions at the observational epoch. To ensure the precision of calibration, we exclude the stars with their renormalised unit weight errors (RUWEs) in \textit{Gaia} catalog larger than 2.0 \citep{Zheng2022AA}. In addition, we calibrate the DCR effect of the matched \textit{Gaia} stars and Alauda according to \citet{Lin2020MNRAS}. Thirdly, we calculate the standard coordinate of each matched star through the central projection \citep{Green1985}, and solve the plate model with a weighted fourth order polynomial \citep{Lin2019MN}, in which some effects such as the geometric distortion and atmospheric refraction can be absorbed. Finally, the observed position (the topocentric astrometric position) of Alauda can be calculated via the solved plate model. Finally, by referring for some ephemeris, the positional $(O-C)$ (the observed minus computed) of Alauda can be conveniently calculated.

\begin{table*}
	\centering
	\small
	\caption{The observations of Alauda overview. The first column shows the observation date. The second column shows the number of frames acquired. The third column shows the exposure time. The fourth column shows the CCD used. The fifth column shows the approximate number of matched \textit{Gaia} stars in the field of view for calibration. The mean seeings of each night are shown in the sixth column and the last column shows the number of close approach events. }
	\begin{tabular}{ccccccc}
		\hline
		\hline
		\makebox[0.13\textwidth][c]{Date (UT)} &  \makebox[0.10\textwidth][c]{Frames (No.)}  & \makebox[0.10\textwidth][c]{ExpTime (s)} & \makebox[0.10\textwidth][c]{CCD used} & \makebox[0.13\textwidth][c]{Calibrated stars (No.)}& \makebox[0.10\textwidth][c]{Seeing ($^\prime$$^\prime$)}  & \makebox[0.13\textwidth][c]{Approach events (No.)}  \\
		\hline
		2020-11-11 &  242     & 40     &  CCD\#1 &  190  & 3.3     & 2       \\
		2020-11-12 &  340     & 30     &  CCD\#1 &  170 &3.2     &  3      \\
		2020-12-08 &  173     & 40     &  CCD\#2 &  860 &1.7     &  4      \\
		2021-01-15 &  130     & 60     &  CCD\#2 &  600 &1.2     &  3       \\
		2021-02-10 &  207     & 60     &  CCD\#2 &  500  &1.2     &   4      \\
		\hline
	\end{tabular}
	\label{table:Observations Details}
\end{table*}

\begin{figure}
	\centering
	\includegraphics[width=0.49\textwidth, angle=0]{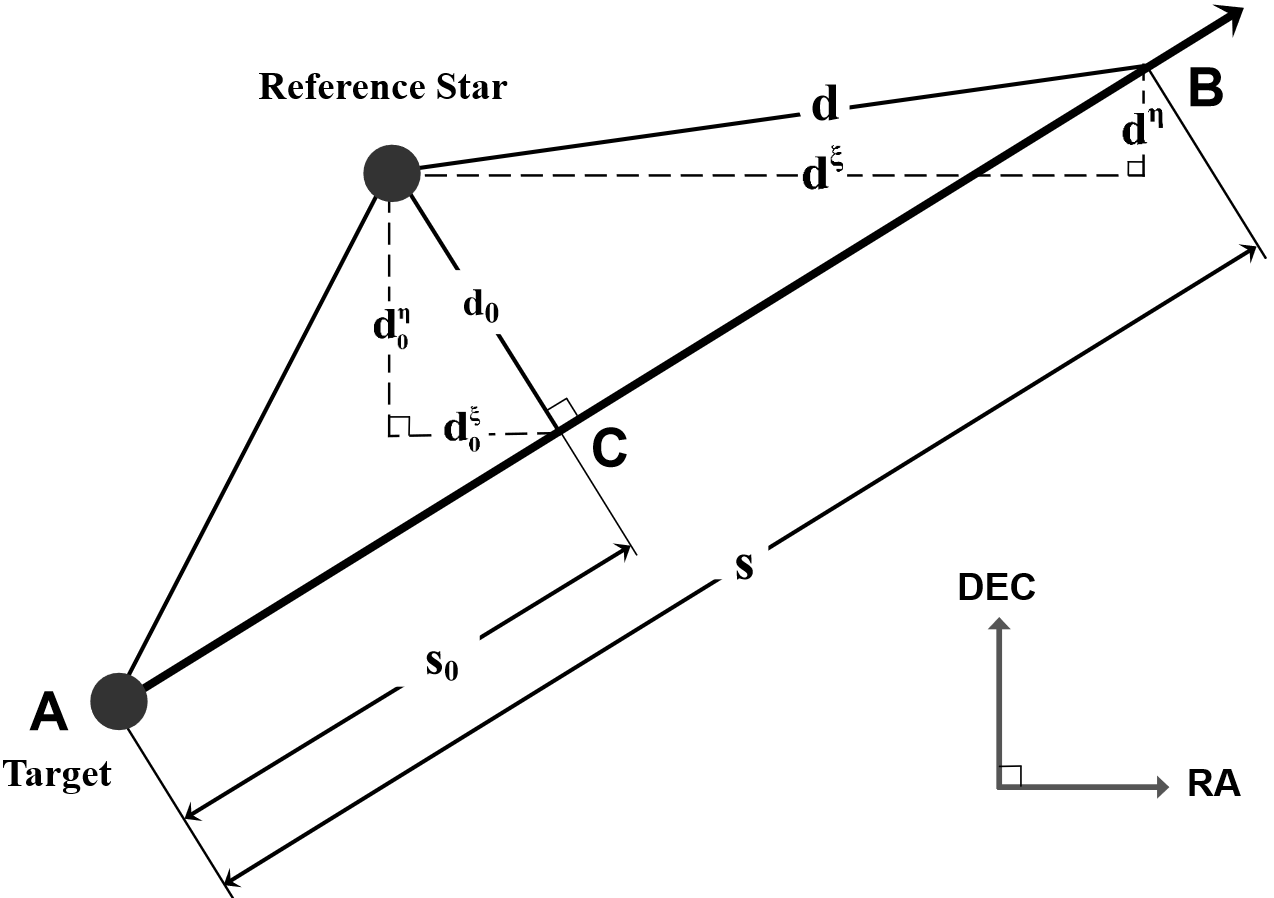}
	\caption{Schematic diagram of the motion model. The target moves relative to a reference star nearby along line $ ACB $ with its
		starting point $ A $ and has its zero recording time ($ t  = 0$) at the point $ A $. The point $ C $ corresponds the relative minimum angular distance $ d_0 $ (also called the impact parameter) between the target and reference star, and has a shift $ s_0 $ from the point $ A $. $ d_0^\xi $ and $ d_0^\eta $ are the projections of $ d_0 $ in right ascension and declination, respectively. At any moment of $ t $, the target moves to the point B with the distance $ d $ relative to the reference star, and the shift $ s $ equals the distance of $ AB $. $ d^\xi $ and $ d^\eta $ are the projections of $ d $ in right ascension and declination, respectively.}
	\label{fig:model}
\end{figure}

\subsection{Close Approach Model}
\label{section:Close Approach Model}

In figure \ref{fig:model}, an asteroid (also called target) moves in the sky plane during a close approach event. It is supposed that the angular velocity can be expressed as a second order polynomial with time in the standard coordinate,
\begin{equation}
	v  = a_0 + a_1t + a_2t^2,
	\label{eq:velocity}
\end{equation}
where $v$ is the angular velocity of the target changing with time, and parameters $ a_0 $ $\sim$ $ a_2 $ are the constants. At any moment of $ t $, the target moves to $ B $ with the shift $ s $ with respect to the starting point $ A $ where its recording time $ t $ is zero. During a close approach event, the asteroid has its maximum approximation at point C. The point $ C $ corresponds the minimum relative angular distance $ d_0 $ (also called the impact parameter) between the target and reference star, and has an offset $ s_0 $ from the point $ A $. At the moment of $ t $, the target has a distance $ d $ relative to the reference star in the standard coordinate. Thus, $ d^2 $ has the following relationship,
\begin{equation}
	d^2 =p_0+p_1t+p_2t^2+p_3t^3+p_4t^4+p_5t^5+p_6t^6.
	\label{eq:squre_d}
\end{equation}
Similar to \citet{Morgado2019MN}, we can fit the square of distance $ d^2 $ in equation (\ref{eq:squre_d}) by the weighted least squares scheme for this sixth order polynomial with 7 parameters ($ p_0 \sim p_6 $). Here, the weighted scheme is used according to the sigmoidal function (precision premium curve) as shown in \citet{Lin2019MN}. The observed positions (which are calculated in Subsection \ref{section:Data Reduction}) of the target and reference star in each frame are used to calculate $ d^2 $ and for fitting. The fitted curve is also called distance curve in \citet{Morgado2019MN}. At the central instant $ t_0 $, the target has its impact parameter $ d_0 $, which can be computed by differentiating equation (\ref{eq:squre_d}) and let it zero. Then, the expression will be shown as the following equation,

\begin{equation}
	p_1+2p_2t_0 + 3p_3t_0^2+4p_4t_0^3+5p_5t_0^4+6p_6t_0^5 = 0,
	\label{eq:differential}
\end{equation}
where $ t_0 $ comes from the determination of the root in equation (\ref{eq:differential}). Based on Newton-Raphson method, we can conveniently solve the value of $ t_0 $ via programming. Then, we can calculate the impact parameter $ d_0 $ ($ d_0 = d $ at the moment of $ t_0 $) according to the solved $ t_0 $ and parameters $ p_0 \sim p_6 $.

\subsection{Calculate the (O -- C) in a Close Approach Event}
\label{section:Calculate OMC}
\par In Figure \ref{fig:model}, the motion of the target in the standard coordinate can be decomposed in right ascension (R.A.) and declination (Decl.). In each direction, we can also respectively fit the square of component distance $ (d^\xi)^2 $ and $ (d^\eta)^2 $ by a sixth order polynomial with time (i.e., equation (\ref{eq:squre_d})). Then, we substitute the central instant $ t_0 $, which corresponds to the resultant minimum distance $ d_0 $ (which is solved in Subsection \ref{section:Close Approach Model}), to the respective fitted sixth order polynomial in each direction. Thus, we can conveniently calculate the components of the minimum distance (impact parameter) $ d_0^{\xi} $ in R.A. and $ d_0^{\eta} $ in Decl. in the standard coordinate. For the positions of the reference star, we can calculate its mean observed position ($ \xi^{R} $, $ \eta^{R} $) (the superscript $ R $ stands for the reference star) in the same standard coordinate. Of course, we can also directly use the computed position based on the catalog (if the catalog provides the precise position of the reference star). Then, the observed position of the target ($ \xi_o^{T} $, $ \eta_o^{T} $) (the superscript $ T $ stands for the target, while the subscript $ O $ stands for the observed position) at the moment of $ t_0 $ can be expressed as,

\begin{equation}
	\left\{
	\begin{aligned}
		\xi_o^{T} & = \xi^{R} + d_0^{\xi} \\
		\eta_o^{T} & = \eta^{R} + d_0^{\eta} \\
	\end{aligned}
	\right. 
	\label{eq:target_rade}
\end{equation}

The signs (positive or negative) of $ d_0^{\xi} $ and $ d_0^{\eta} $ should be determined according to the relative position between the target and reference star by the observation near the moment of $ t_0 $. Through re-projection \citep{Green1985}, we can calculate the observed position of the target ($ \alpha_o^{T} $, $ \delta_o^{T} $) in equatorial coordinate. We can acquire the computed position of the target ($ \alpha_c^{T} $, $ \delta_c^{T} $) from some ephemeris by submitting the central instant $ t_0 $. Finally, we can calculate the positional $ (O-C) $ (the observed minus computed) of the target in a close approach event.

\begin{table*}
	\centering
	\scriptsize
	\caption{The reduced results for all the used observations of main-belt asteroid (702) Alauda. The second column shows the \textit{Gaia} IDs of the reference stars for each close approach event. The third column gives the central instants ($ t_0 $) in form of JD-2459000. The fourth and fifth columns list the $ (O-C) $s of Alauda in R.A. and Decl., which are derived from the fitted scheme of both 4th order and 6th order polynomials, respectively. The impact parameters and  magnitudes of the reference stars (\textit{Gaia-G} magnitude) are listed in the sixth and seventh columns, respectively. The eighth column shows the estimated uncertainty of central instant according to \citet{Morgado2016MN}. The ninth and tenth columns respectively show the observed minimum and maximum angular distance between Alauda and the reference star. The last column shows the duration for each close approach event. Below the results of each observation set, the mean positional $ (O-C) $s and standard deviations are also listed together with conventional solutions.}
	\begin{tabular}{ccccccccccc}
		\hline
		\hline
		\makebox[0.04\textwidth][c]{Date (UT)} &  \makebox[0.08\textwidth][c]{Reference Star ID}  & \makebox[0.09\textwidth][c]{$ t_0 $ (JD-2459000)} & \makebox[0.10\textwidth][c]{$\Delta\alpha\cos\delta$ ($^\prime$$^\prime$)} & \makebox[0.10\textwidth][c]{$\Delta\delta$ ($^\prime$$^\prime$)}  & \makebox[0.05\textwidth][c]{$ d_0 $} & \makebox[0.04\textwidth][c]{Star Mag} &\makebox[0.04\textwidth][c]{$\sigma_{t_{0}}$} &  \makebox[0.04\textwidth][c]{Min Dis }& \makebox[0.04\textwidth][c]{Max Dis }
		&  \makebox[0.04\textwidth][c]{Duration} \\
		&        &       &  6th order (4th order)   &   6th order (4th order)     &  ($^\prime$$^\prime$)   &  & (s) &($^\prime$$^\prime$)& ($^\prime$$^\prime$) &(hour)\\
		\hline
		2020-11-11 &  190965107955867776     & 165.3385796     &  -0.003 (-0.003)  &    \ 0.001 (-0.002)    & \ \ 7.81    &14.2 & 0.784  &\ \ 7.81     & 60.85   & 2.80\\
		2020-11-11 &  190965103659693952     & 165.3713975     &  \ 0.003 (\ 0.006)  &     -0.001 (-0.002)     &  14.68  &13.8  &0.938  &14.68   & 36.09  & 2.80   \\
		2020-11-12 &  190945110588179456     & 166.3279130     &  \ 0.004 (\ 0.007) &    -0.004 (-0.002)    &  33.09    &15.3 & 1.682 &  33.08 &  57.61    & 3.05   \\
		2020-11-12 &  190945106293176064     & 166.3411577     &  \ 0.000 (-0.001) &    -0.004 (-0.003)     &  19.22   &16.2 & 3.125 & 19.23 & 44.57      &3.05   \\
		2020-11-12 &  190945316746612608     & 166.3598021     &  \ 0.004 (-0.007) &    -0.001 (-0.001)     &  24.57  &15.4 & 1.282  & 24.58 & 41.48       &3.05  \\
		\hline
		& & Average position (4th order)          &   0.001 $\pm$ 0.006        &    -0.002 $\pm$ 0.001                    & &      &  && \\
		& & Average position (6th order)          &   0.001 $\pm$ 0.003        &    -0.002 $\pm$ 0.002                    & &      &   &&\\
		& & Conventional solution &   0.006 $\pm$ 0.010        &    -0.002 $\pm$ 0.008                     & &         &  &&\\
		\hline
		2020-12-08 &  188197534110556032     & 192.1526737     &  \ 0.004 (\ 0.005) &    \ 0.004 (\ 0.002)    &  30.94&16.0&2.322  &30.93 & 70.54 &2.44 \\
		2020-12-08 &  188197534110557952     & 192.1595573     &  -0.002 (-0.009) &    -0.002  (-0.001)    &  23.43  &15.0 & 0.988 & 23.42 & 62.34 &2.44 \\
		2020-12-08 &  188197534110557696     & 192.1675330     &  \ 0.004 (\ 0.002) &    -0.002  (-0.001)   &  38.02 &15.4 & 1.422  & 38.02 & 63.85 &2.44 \\
		\hline
		& & Average position (4th order)          &   -0.002 $\pm$ 0.007        &    \ 0.000 $\pm$ 0.002                    & &       &   && \\
		& & Average position (6th order)          &   \ 0.002 $\pm$ 0.003        &    \ 0.000 $\pm$ 0.003                    & &       &  && \\
		& & Conventional solution    &   -0.001 $\pm$ 0.012        &    -0.001 $\pm$ 0.008                     & &       &    && \\
		\hline
		2021-01-15 &  173523550707799552     & 230.0214632     &  -0.003 (\ 0.002) &    -0.011 (-0.006)    &  15.17  &16.7&1.454  & 15.15 & 43.50  &2.54 \\
		2021-01-15 &  173523275829562624     & 230.0590659     &  -0.005 (-0.008) &    -0.014 (-0.006)    &  72.65   &13.8 &2.245 &  72.65 & 82.44 &2.54 \\
		2021-01-15 &  173523344549362432     & 230.0668542     & \  0.000 (-0.003)  &   -0.006 (-0.005)    &  32.20  &15.6 &1.355  & 32.21 & 54.01 &2.54 \\
		\hline
		& & Average position (4th order)         &   -0.003 $\pm$ 0.005        &    -0.006 $\pm$ 0.001                    & &      &   && \\
		& & Average position (6th order)         &   -0.003 $\pm$ 0.003        &    -0.010 $\pm$ 0.004                    & &      &   && \\
		& & Conventional solution   &   -0.001 $\pm$ 0.007        &    -0.005 $\pm$ 0.009                     & &      &   &&   \\
		\hline
		2021-02-10 &  159439940627025408     & 256.0220522     &  \ 0.000 (\ 0.003) &    \ 0.001 (\ 0.000)     &  47.25 &12.1 &3.426   &  47.24&76.93 &4.27  \\
		2021-02-10 &  159440048001339008     & 256.0370666     &  -0.001 (\ 0.000) &     -0.002 (-0.002)     &  27.49 &15.4 &1.720   &  27.48&60.72  &4.27 \\
		2021-02-10 &  159439910562389760     & 256.0984256     &  \ 0.000 (\ 0.004) &    -0.003 (-0.003)     &  53.03 &14.7 &1.832   &  53.03& 69.98  & 4.27 \\
		2021-02-10 &  159439738763696640     & 256.1304893     &  -0.002 (-0.001) &    \ 0.000  (-0.003)     &  25.45  &16.0 &2.162  & 25.44 &64.39  &4.27 \\
		\hline
		& & Average position (4th order)          &   \ 0.001 $\pm$ 0.002        &    -0.002 $\pm$ 0.002                    & &      &    &&  \\
		& & Average position (6th order)          &   -0.001 $\pm$ 0.001        &    -0.001 $\pm$ 0.001                    & &        &  &&  \\
		& & Conventional solution    &   \ 0.001 $\pm$ 0.007        &    -0.003 $\pm$ 0.010                     & &      &   && \\
		\hline
	\end{tabular}
	\label{table:Results_astrometry}
\end{table*}

\begin{figure*}
	\includegraphics[width=0.95\textwidth]{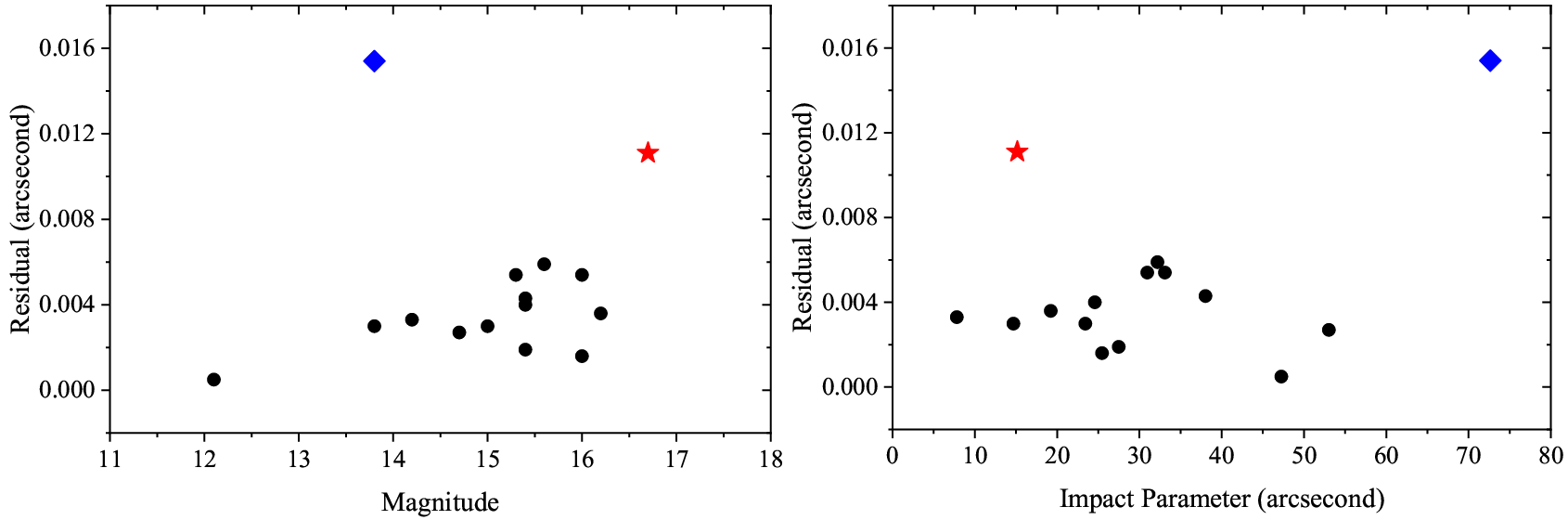}
    \caption{In this figure, the left panel shows the distributions of the absolute positional residuals ($ \sqrt{{(\Delta\alpha\cos\delta)}^2+{(\Delta\delta)}^2} $) derived from close approach events against \textit{Gaia-G} magnitudes, while the right panel shows the distributions against impact parameters. The residuals plotted in blue diamond and red asterisk are distinguished for their larger values. The residual in blue diamond is large (left panel) might be due to its large impact factor (right panel). The residual in red asterisk is large (right panel) might be due to the faintness of the reference star (left panel).}
	\label{fig:relationship}
\end{figure*}

\subsection{Comparison with Previous Works}
\par Although our reduction method of close approach event is somewhat similar to \citet{Morgado2016MN,Morgado2019MN}, there are still some differences.  The major difference is the coordinate system used during the reduction process. They reduced the observations in pixel coordinate, which introduces some problems: the determination of pixel scale, the orientation of CCD and geometric distortion. In this work, we reduce the observations in standard coordinate (the projection of equatorial coordinate), in which these effects can be calibrated in the plate model. In addition, we extend the method from two Galilean moons to a main-belt asteroid and its nearby reference star. \citet{Morgado2016MN,Morgado2019MN} supposed the angular velocity with a first order polynomial with time and fitted the distance curve (equation (\ref{eq:squre_d})) with a fourth order polynomial, while we supposed the angular velocity with a second order polynomial and fit the distance curve with a sixth order polynomial to better reflect the accuracy and precision in the practice of a main-belt asteroid, which will be further discussed in Section \ref{section: Results}.

\section{Results}
\label{section: Results}
\subsection{Astrometric Results}
\label{section: Astrometric Results}
\par As shown in Subsection \ref{section:Data Reduction}, we calculate the positional $ (O-C) $s of Alauda in each frame, which are here called the conventional solutions. The observed positions (as calculated in Subsection \ref{section:Data Reduction}) in each frame are further used for calculation and fitting to derive the observed positions of each close approach event. For this astrometric reduction, the positional effects of geometric distortion, atmospheric refraction, DCR effect have been taken into account. In addition, we notice that there is a satellite of Alauda, which might affect our positional measurement. According to light ratio, mass ratio and the angular distance between the primary and its satellite presented in \citet{Rojo2011ApJ}, the evaluated positional difference between their photocenter and the center of mass caused by the satellite is less than negligible 1 milli-arcsecond using the method of \citet{Peng2023AA}.

\par In order to derive the positional $ (O-C) $ of Alauda for each close approach event (at the moment of maximum approximation), we compute the square of the observed distance $ d^2 $ between the target and reference star in the standard coordinate. Then, we can solve the central instant $ t_0 $ by solving the root of equation (\ref{eq:differential}). Meanwhile, we also fit the six order polynomials (equation (\ref{eq:squre_d})) in the direction of R.A. and Decl.. Based on the solved central instant $ t_0 $, we can calculate the component of the minimum distance $ d_0^{\xi} $ in R.A. and $ d_0^{\eta} $ in Decl.. According to equation (\ref{eq:target_rade}), we calculate the observed positions of the target and project it to the equatorial coordinate. The computed positions of the target are derived from JPL ephemeris\footnote{https://ssd.jpl.nasa.gov/horizons.cgi\#top} (JPL\#116 and DE441). For comparison, we also calculate the results if supposing the squares of the relative distances ($ d^2 $, $ (d^\xi)^2 $, $ (d^\eta)^2 $) are the fourth order polynomials with time. The results of the positional $ (O-C) $ are shown in Table \ref{table:Results_astrometry} together with some information of each close approach event.

\begin{figure*}
	\includegraphics[width=0.9\textwidth]{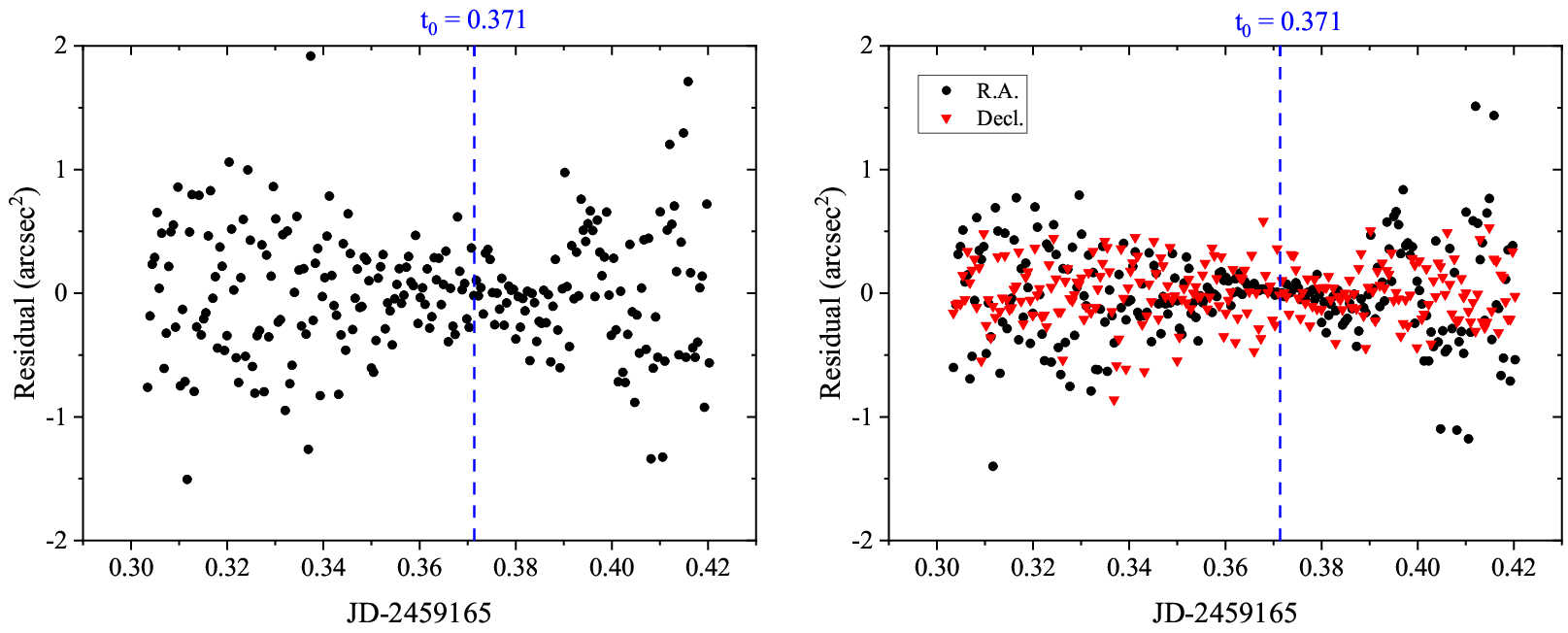}
	\caption{The fitted residuals of $ d^2 $ in equation (\ref{eq:squre_d}) with time in a close approach event observed on Nov 11, 2020. The \textit{Gaia} ID of the reference star is 190965103659693952. The left panel shows the resultant fitted residuals, while the right panel shows the components in the direction of R.A. and Decl..}
	\label{fig:20201111-3952}
\end{figure*}

\par  Compared with conventional solutions, the positional precision of results from close approach events (fitted with both fourth and sixth order polynomial) are improved significantly. The observed topocentric astrometric positions of Alauda (fitted by six order polynomials) via close approach events are shown in Table \ref{table:topocentric astrometric positions}. For the scheme of fitted by a fourth order polynomial, the standard variations of each observation set are 0\farcs002 $\sim$ 0\farcs007 and 0\farcs001 $\sim$ 0\farcs002 in R.A. and Decl., respectively. For the scheme of fitted by a sixth order polynomial, the standard variations of each observation set are 0\farcs001 $\sim$ 0\farcs003 and 0\farcs001 $\sim$ 0\farcs004 in R.A. and Decl., respectively. Both the mean $ (O-C) $s from close approach events are similar to the results of conventional solutions. In our observations, the angular velocity of the target in R.A. usually strides across the peak of the velocity variation caused by diurnal parallax, while the angular velocity in Decl. is not. Considering the effect of diurnal parallax, we adopt the scheme of fitting by a sixth order polynomial, which will better reflect the accuracy and precision in the practice. Usually for a main-belt asteroid, if we observe the target within a short time span (e.g. within 2 hours) and the angular velocity will not strides across the peak of the velocity variation, a fourth order polynomial for fitted can derive good results. In Subsection \ref{section: Discussion}, we will present the difference of the fitted residual in a specific case.

\begin{table}
	\centering
	\small
	\caption{The table shows the topocentric astrometric positions of asteroid (702) Alauda via close approach event. The IAU code of the observatory is 286. The first column lists the Julian dates of maximum approximation. The second and third columns list the corresponding positions in right ascension (R.A.) and declination (Decl.), respectively.}
	\begin{tabular}{ccc}
		\hline
		\hline
		\makebox[0.15\textwidth][c]{Julian Date} &\makebox[0.12\textwidth][c]{R.A. (h m s)}& \makebox[0.12\textwidth][c]{Decl. ($^{\circ}$ $^{\prime}$ $^{\prime\prime}$)}  \\
		\hline
		2459165.3385796 & 05 36 42.5223 & +39 46 37.745 \\
		2459165.3713975 & 05 36 41.1470 & +39 46 35.894 \\
		2459166.3279130 & 05 36 02.0518 & +39 45 42.400 \\
		2459166.3411577 & 05 36 01.4756 & +39 45 41.572 \\
		2459166.3598021 & 05 36 00.6663 & +39 45 40.382 \\
		2459192.1526737 & 05 12 10.8423 & +38 29 09.676 \\
		2459192.1595573 & 05 12 10.3978 & +38 29 07.612 \\
		2459192.1675330 & 05 12 09.8837 & +38 29 05.222 \\
		2459230.0214632 & 04 41 41.7784 & +34 02 06.951 \\
		2459230.0590659 & 04 41 40.8386 & +34 01 49.600 \\
		2459230.0668542 & 04 41 40.6437 & +34 01 45.992 \\
		2459256.0220522 & 04 40 31.0515 & +31 00 06.122 \\
		2459256.0370666 & 04 40 31.2851 & +31 00 00.680 \\
		2459256.0984256 & 04 40 32.2478 & +30 59 38.324 \\
		2459256.1304893 & 04 40 32.7584 & +30 59 26.573 \\
		\hline
	\end{tabular}
	\label{table:topocentric astrometric positions}
\end{table}

\par With the fitted scheme of a six order polynomial, we do not find the obvious relevance between positional accuracy and magnitudes of reference stars or the impact parameters, which can be shown in Figure \ref{fig:relationship}, because the effects of atmospheric turbulence and local telescope optics might not be dominant in positional errors. The positional accuracy of each event might be also affected by the other conditions such as the uncertainty of central instant. However, we think the dominating effect in this work is the centering errors of Alauda and reference stars, which will be elaborated in the following subsection. In Figure \ref{fig:relationship}, most of the absolute positional residuals ($ \sqrt{{(\Delta\alpha\cos\delta)}^2+{(\Delta\delta)}^2} $) range from 2 to 6 milli-arcseconds, except two in different shape. The positional absolute residual in blue diamond might be affected by the larger impact parameter, while the residual in red asterisk might be due to the faintness of the reference star.

\subsection{Case Analysis}
\par In this subsection, we will analyze two typical cases for the close approach events in this work. For more details of the close approach events, one can find the figures of all the distance curves and the fitted residuals with a sixth order polynomial in the Appendix.
\par Figure \ref{fig:20201111-3952} shows the fitted residuals of equation (\ref{eq:squre_d}) with time when Alauda approached the reference star whose \textit{Gaia} ID is 190965103659693952 observed on Nov 11, 2020. The left panel shows the fitted residuals of resultant $ d^2 $, while the right panel shows residuals of the components $ (d^{\xi})^2 $ in R.A. and $ (d^{\eta})^2$ in Decl.. The residual dispersion of $ d^2 $, $ (d^{\xi})^2 $ and $ (d^{\eta})^2$ almost show the consistency with time. Near the central instant $ t_0 = 0.371$ (with the minimum relative angular distance $ d_0 $), the dispersion of fitted residuals reach the minimum, which is mainly due to the similar effects of atmospheric turbulence and local telescope optics when the relative angular distance is small. However, the minimum dispersion may not be exactly located at the moment of $ t_0 $ for the possible reasons of the change of seeing, air mass and the characteristics of local CCD chip. The mentioned factors affect the precision of centering, and further affect the observed $ d^2 $ for fitting.

\par
\begin{figure*}
	\includegraphics[width=0.9\textwidth]{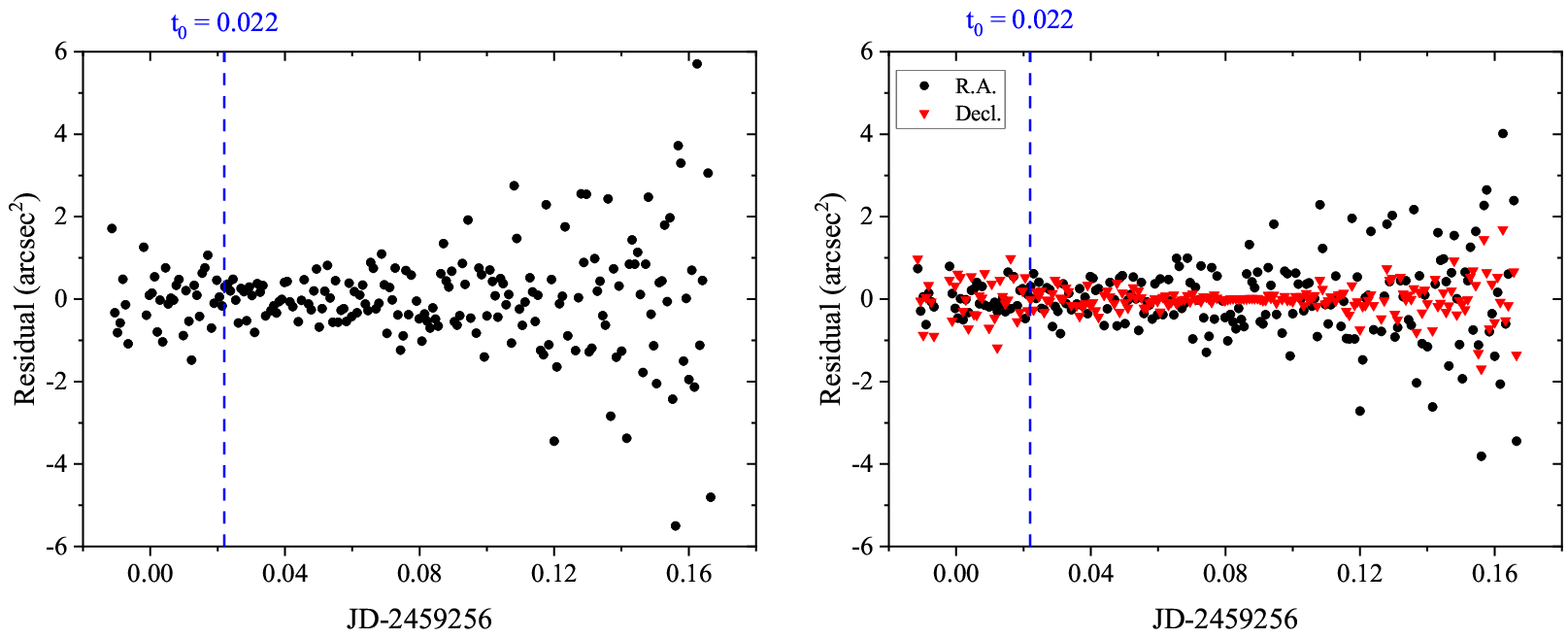}
	\caption{The fitted residuals of $ d^2 $ in equation (\ref{eq:squre_d}) with time in a close approach event observed on Feb 10, 2021. The \textit{Gaia} ID of the reference star is 159439940627025408. The left panel shows the resultant fitted residuals, while the right panel shows the components in the direction of R.A. and Decl..}
	\label{fig:20210210-5408}
\end{figure*}

\begin{figure}
	\includegraphics[width=0.45\textwidth]{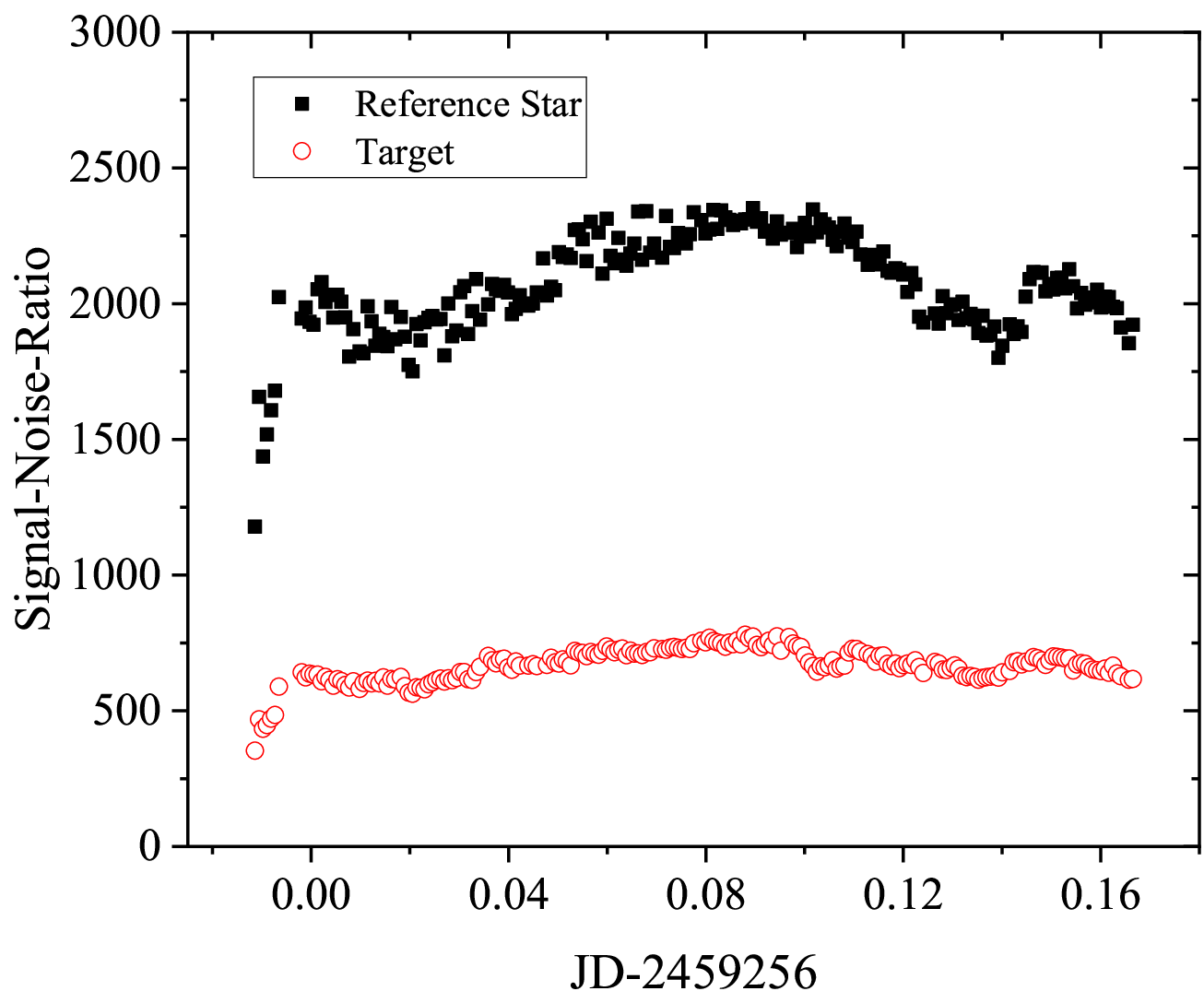}
	\caption{The figure shows the change of SNRs of Alauda and reference star  (\textit{Gaia} ID: 159439940627025408) with time on Feb 10, 2021. The red and black spots shows the SNRs of Alauda and the reference star, respectively.}
	\label{fig:SNR}
\end{figure}

 Figure \ref{fig:20210210-5408} shows the fitted residuals of $ d^2 $ in equation (\ref{eq:squre_d}) for another close approach event observed on Feb 10, 2021 with the nearby reference star's \textit{Gaia} ID 159439940627025408. For the left panel, we notice that near the moment of $ t_0 = 0.022$, the dispersion of fitted residual for $ d^2 $ is small. The dispersion reaches the minimum near the moment of $ t = 0.035 $. From the right panel, the dispersion of $ (d^{\xi})^2 $ in R.A. reaches the minimum near the moment of $ t = 0.035 $, whereas the dispersion $ (d^{\eta})^2 $ in Decl. reaches the minimum near the moment of $ t = 0.08 $. The reasons are presented as follows. Firstly, near the moment of $ t = 0.08 $, the SNRs of the sources in the field of view are higher due to the better weather conditions, and the centering errors of both Alauda and the reference star become smaller. Figure \ref{fig:SNR} shows the change of SNRs of both Alauda and the reference star. It can be seen that near the moment of $ t = 0.08 $, the SNRs of the reference star are larger due to the  better weather conditions, which makes the positional precision better. The SNRs of Alauda will be affected by the weather conditions, other nearby stars and its rotation. Secondly, the telescope used is equatorial-mounted. The tracking accuracy in R.A. becomes worse when pointing to the west. Hence, the dispersion residuals in the direction of R.A. and Decl. are not consistent.

\begin{figure*}
	\includegraphics[width=0.9\textwidth]{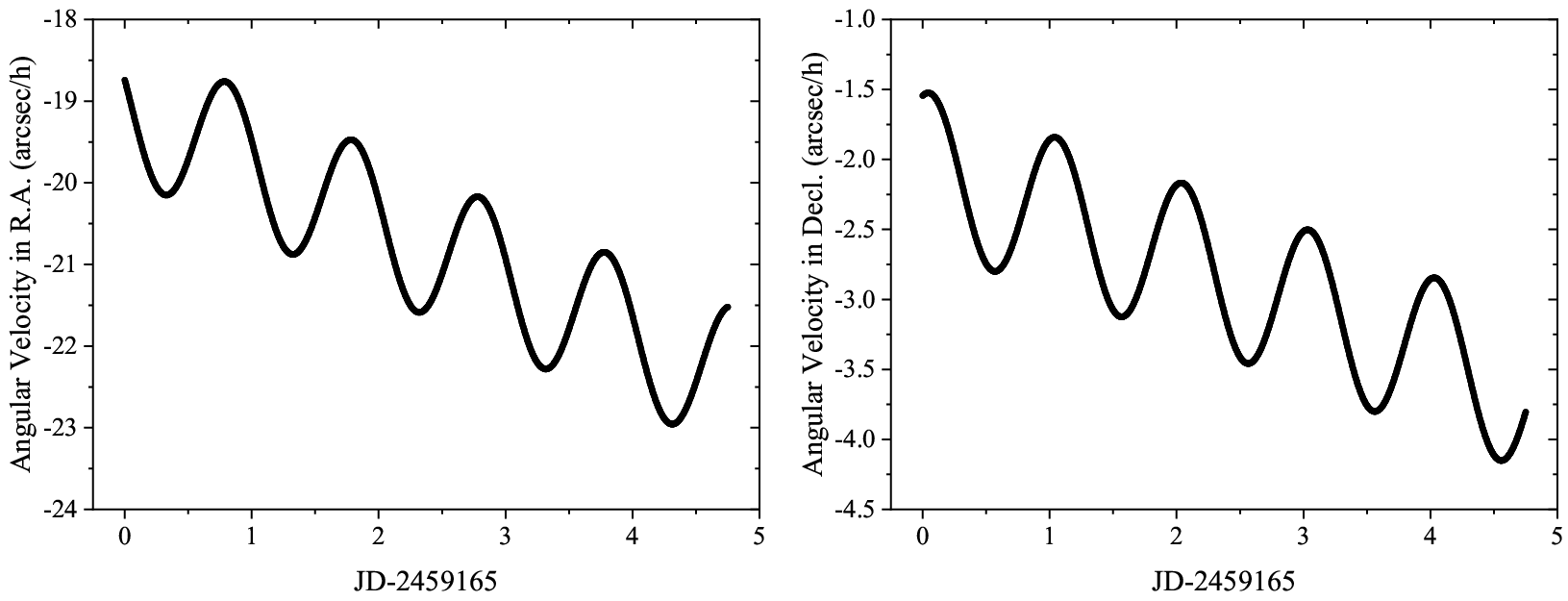}
	\caption{The change of Alauda's angular velocity with time of JD 2459165 $\sim$ 2459169 (Nov 11 $\sim$ 15, 2020) from JPL ephemeris. The left and right panels show the angular velocity with time in R.A and Decl., respectively.}
	\label{fig:velocity_radec}
\end{figure*}

\begin{figure}
	\centering
	\includegraphics[width=0.45\textwidth]{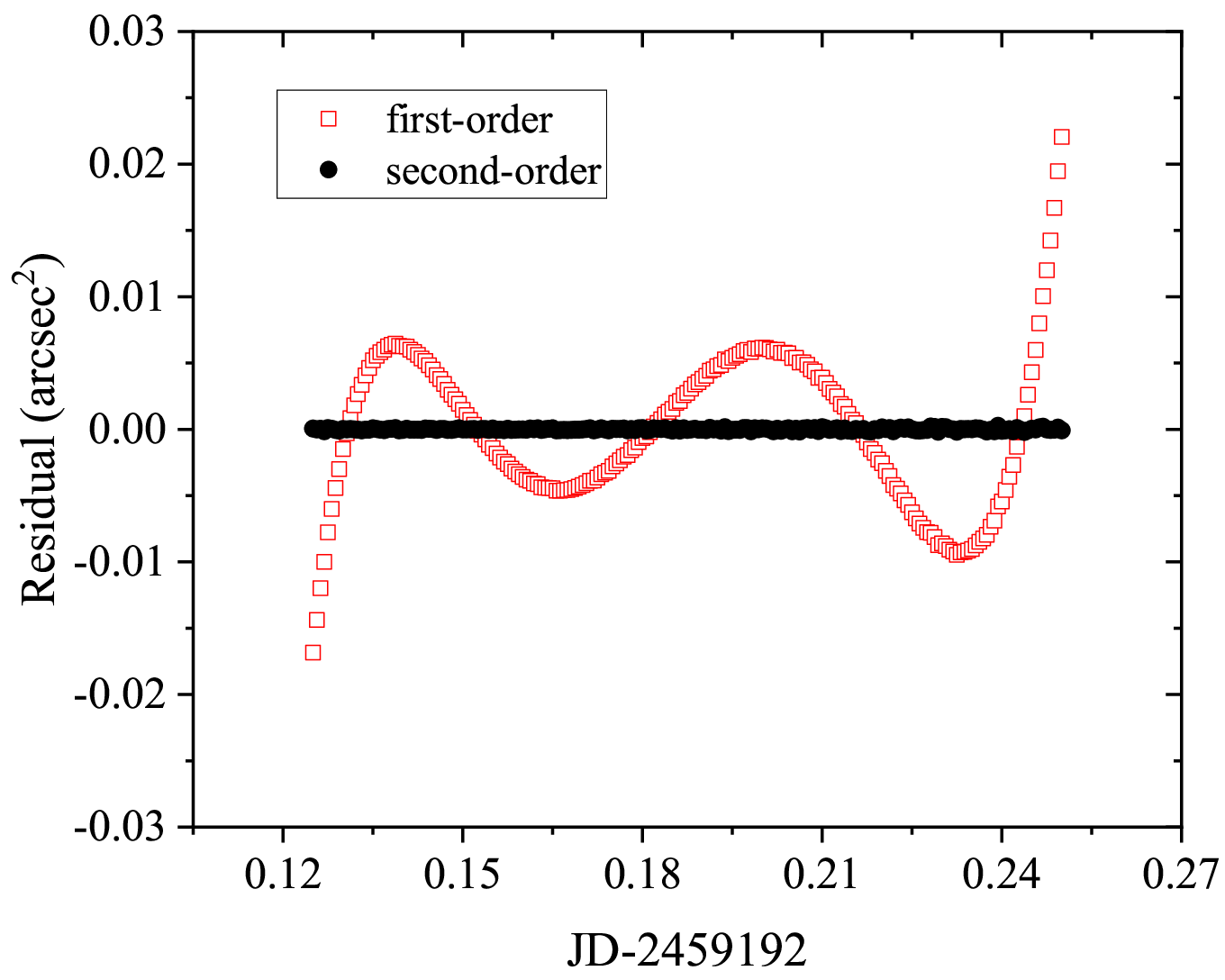}
	\caption{The figure shows the fitted residuals of $ d^2 $ with time. The black plots shows the fitted residuals with the angular velocity model with a first order polynomial, while the red shows fitted residuals of the angular velocity model with a second order polynomial.}
	\label{fig:fitted_compare}
\end{figure}

\par Theoretically, in a close approach event, the dispersion of fitted residual of $ d^2 $ will reach the minimum with the minimum relative angular distance $ d_0 $ at the moment of $ t_0 $. However, in fact, the centering errors will be affected by some factors such as the change of weather conditions, air mass, tracking accuracy, dome occlusion and local characteristics of CCD chip. These effects might affect and dominate the positional errors. If the asteroid with higher SNR approaches brighter reference stars, the positional errors might be dominated by the atmospheric turbulence and local telescope optics. In such a case, we believe that more precise positions will be obtained from close approach events.

\subsection{Discussion}
\label{section: Discussion}
\par In this work, we suppose that the angular velocity of a main-belt asteroid is a second order polynomial with time as shown in equation (\ref{eq:velocity}) in a close approach event. Figure \ref{fig:velocity_radec} shows the change of angular velocity with time from Nov 11 to 15, 2021 in both R.A. and Decl. from JPL ephemeris. It can be seen that the angular velocity changes sinusoidally with time due to the diurnal parallax. During the time, the declination of Alauda is about 40$^\circ$ and the amplitudes in both R.A. and Decl. are numerically similar. For the asteroids (especially for NEAs) in a close approach event, to obtain the precise positions with the level of several milli-arcseconds, the angular velocity model of at least the second order polynomial with time should be supposed. For comparison, Figure \ref{fig:fitted_compare} shows the fitted residuals of $ d^2 $ with different supposed angular velocity models in the close approach event of reference star ID 188197534110557952 on Dec 8, 2020. The positions of Alauda are taken from JPL ephemeris while the position of the star are from \textit{Gaia} DR3 catalog. The angular velocity model with a second order polynomial ($ d^2 $ is a sixth order polynomial with time) can be fitted better with the residuals close to zero. If we take the angular velocity model with a first order polynomial ($ d^2 $ is a fourth order polynomial with time), the fitted residuals can reach about 0.2 $ arcsec^2  $ (depend on the central instant), which might cause the positional errors of several milli-arcseconds in a close approach event. With the fixed fitted errors, the positional errors depend on the values of impact parameter ($ d_0 $, $ d_0^{\xi} $ and $ d_0^{\eta} $). The smaller the impact parameter is, the larger positional error will cause. The difference of the two fitted velocity models (first-order or second-order polynomial) is the consideration of diurnal parallax effect. We can fit the diurnal parallax effect well using a second-order polynomial velocity model. There will be larger diurnal parallax effect for angular velocities of NEAs. Therefore, we should also consider this effect using the second-order polynomial velocity model for NEAs to derive accurate positions. For Kuiper-belt objects (KBOs), both their angular velocities and diurnal parallax effects are much smaller. The technique in this work might not work well for KBOs because they only pass across the small angular distance during one night. However, if one would like to make use of the close approach event, we still recommend using the second-order polynomial velocity model so that the diurnal parallax effect can be well considered. Perhaps, for the close approach event of natural satellites, the angular velocity model with higher order polynomial with time should be supposed to obtain the precise positions (e.g., with the precision better than 5 milli-arcseconds).

\par The keys to obtaining precise position in this work are presented as follows. Firstly, we reduce the observations in standard coordinate instead of pixel coordinate. In this way,  the effects of pixel scale, CCD orientation, geometric distortion and atmospheric refraction can be taken into account using accurate and precise positions of calibrated stars provided by \textit{Gaia} catalog. Secondly, the DCR effect of both the target and reference star should be calibrated. The positional systematical errors affect the calculation of the relative angular distance between the target and reference star. Even if we used the \textit{Cousins-I} filter to minimize the DCR effect, the positional errors of relative angular distance might reach about 10 milli-arcseconds for each frame (mainly depend on the observed zenith distance and the different color indices between the target and reference stars). Thirdly, when the object passes through a dense star field, observations of multiple close approach events can be captured during one night and more precise positions can be obtained. To fully tap the potential of close approach event, the target and nearby reference star should be bright enough so that the effects of atmospheric turbulence and local telescope optics dominate the positional errors.

\section{Conclusion}
\label{section:Conclusion}

\par In this work, we explore a method to obtain the precise positions of a main-belt asteroid in the close approach event. We perform reduction in the standard coordinate to calibrate the effects of pixel scale, CCD orientation and geometric distortion. To further improve the fitted precision, the angular velocity model with a second order polynomial is supposed. In practice, 15 close approach events of the main-belt asteroid (702) Alauda are taken over 5 nights. The positional precision of the close approach events reach 1$\sim$3 and 1$\sim$4 milli-arcseconds in R.A., and Decl., respectively. Compared with the similar work of \citet{Morgado2019MN}, we extend the application from Galilean moons to a main-belt asteroid and the reference star nearby. We derive better precision for the asteroid (702) Alauda compared with the precision of 11.3 milli-arcseconds in the practice of Galilean moons. If we tap the potential of the close approach event in the proper conditions, the positional precision might be comparable to that is from stellar occultation.

\section*{Acknowledgements}

This work was supported by the National Key R\&D Program of China (Grant No. 2022YFE0116800), by the China Manned Space Project (Grant No. CMS-CSST-2021-B08), by the National Natural Science Foundation of China (Grant Nos. 11873026, 11273014) and Joint Research Fund in Astronomy (Grant No. U1431227). We thank the anonymous reviewer who provided us with valuable comments, Dr. Lin F. R. who had a helpful discussion with us and Mr. Cao J. L. who helped us match the observations. This work has made use of data from the European Space Agency (ESA) mission Gaia (\url {https://www. cosmos.esa.int/gaia}), processed by the Gaia Data Processing and Analysis Consortium (DPAC; \url{hppts://www.cosmos.esa.int /web/gaia/dpac/consortium}). Funding for the DPAC has been provided by national institutions, in particular the institutions participating in the Gaia Multilateral Agreement.

\section*{Data Availability}

The data underlying this article will be shared on reasonable request to the corresponding author.



\bibliographystyle{mnras}
\bibliography{example} 




\appendix

\section{Distance Curve and Fitted Residual}

\begin{figure}
	\centering
	\includegraphics[width=0.46\textwidth, angle=0]{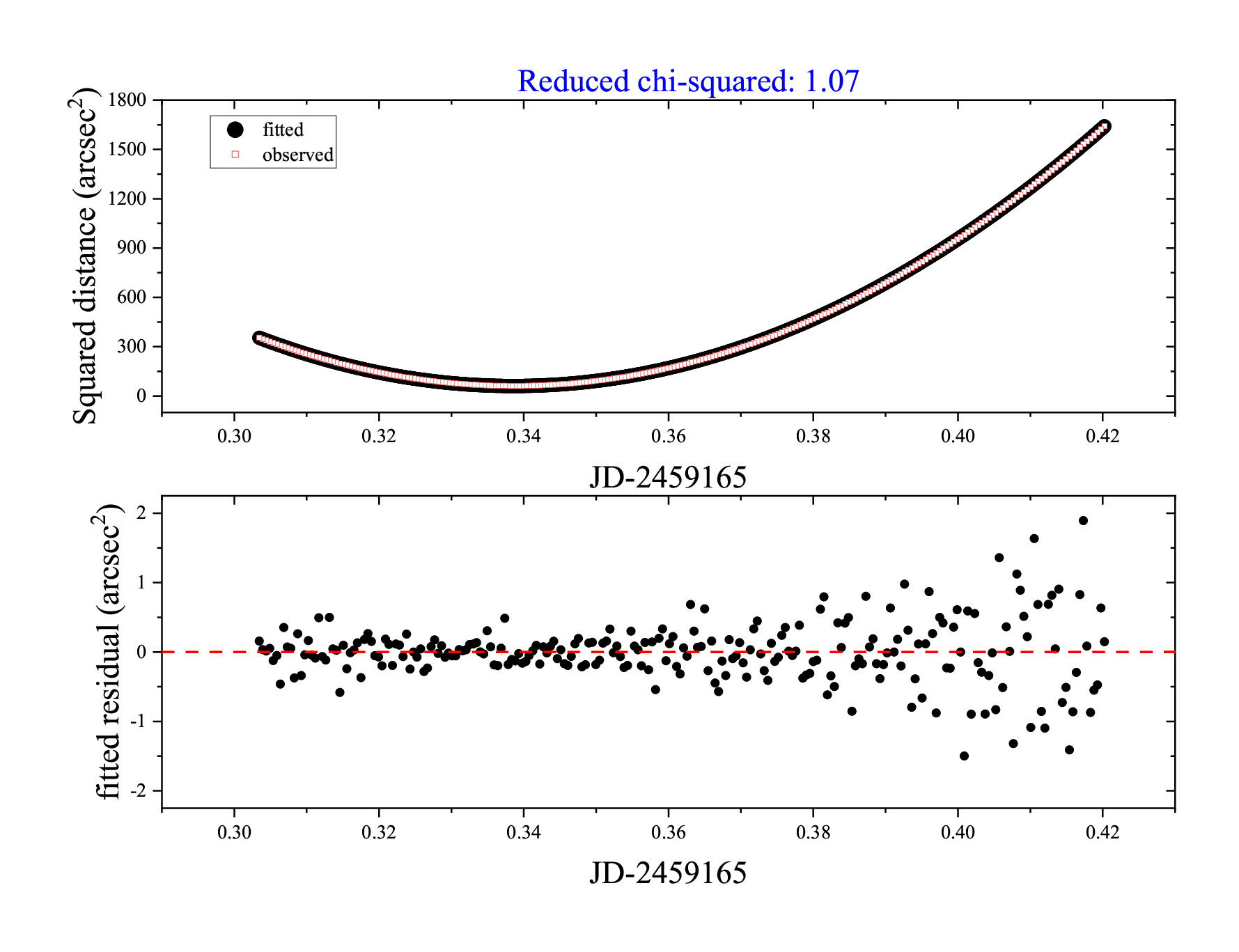}
	\caption{The figure shows the distance curve and fitted residuals of $ d^2 $ with time. The reference star ID is 190965107955867776, approached on Nov 11, 2020.}
	\label{fig:final_7776}
\end{figure}

\begin{figure}
	\centering
	\includegraphics[width=0.45\textwidth, angle=0]{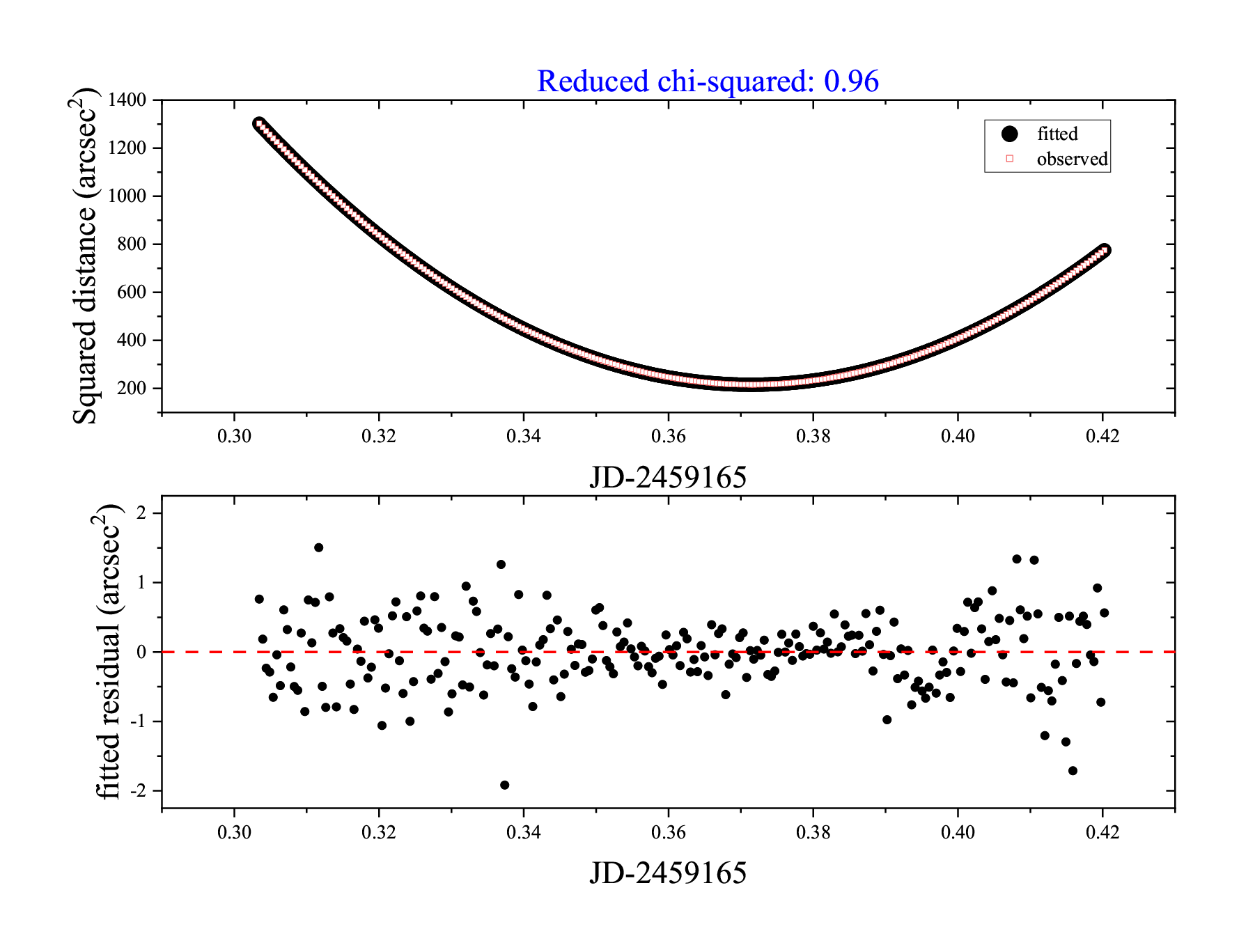}
	\caption{The figure shows the distance curve and fitted residuals of $ d^2 $ with time. The reference star ID is 190965103659693952, approached on Nov 11, 2020.}
	\label{fig:final_3952}
\end{figure}

\begin{figure}
	\centering
	\includegraphics[width=0.45\textwidth, angle=0]{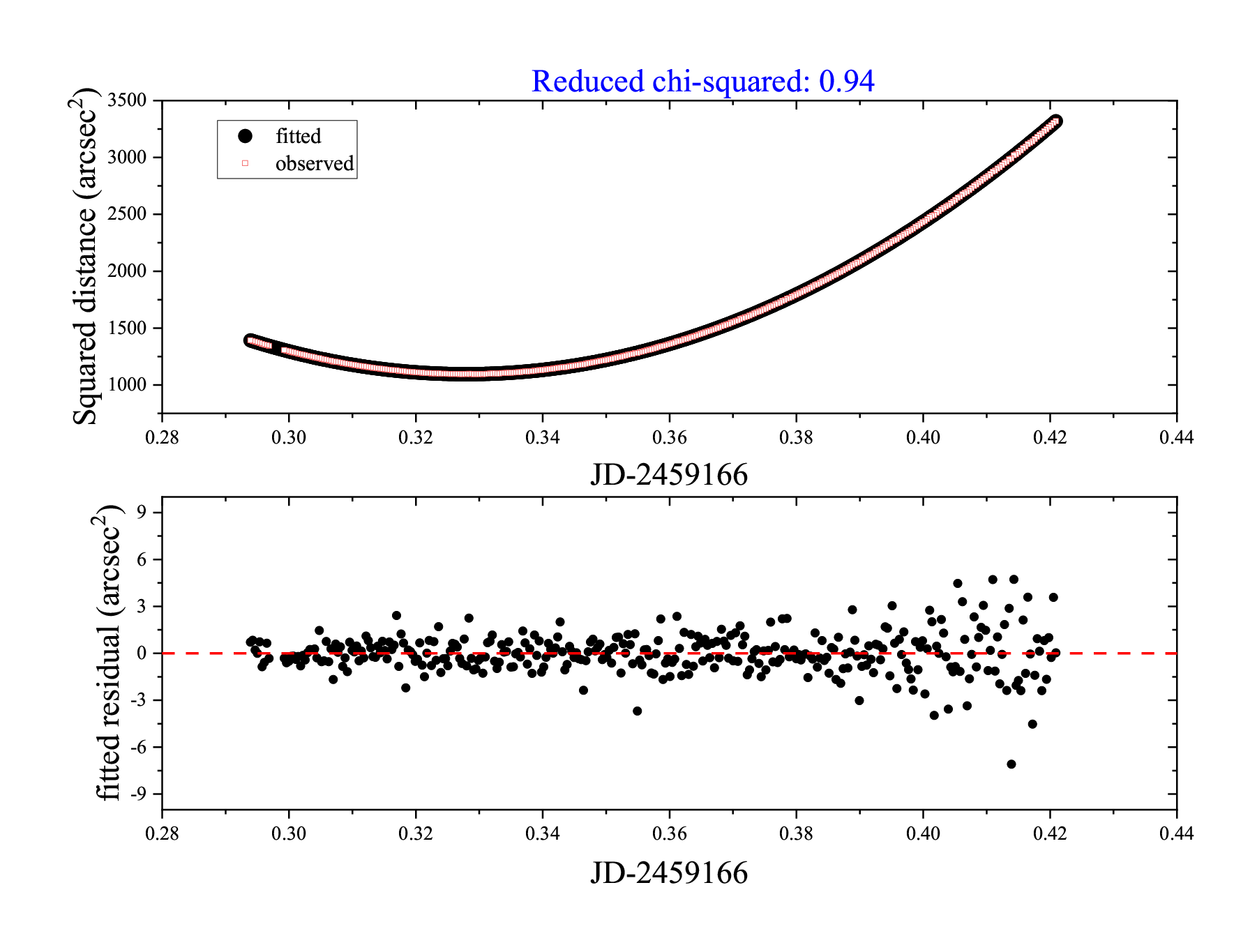}
	\caption{The figure shows the distance curve and fitted residuals of $ d^2 $ with time. The reference star ID is 190945110588179456, approached on Nov 12, 2020.}
	\label{fig:final_9456}
\end{figure}

\begin{figure}
	\centering
	\includegraphics[width=0.46\textwidth, angle=0]{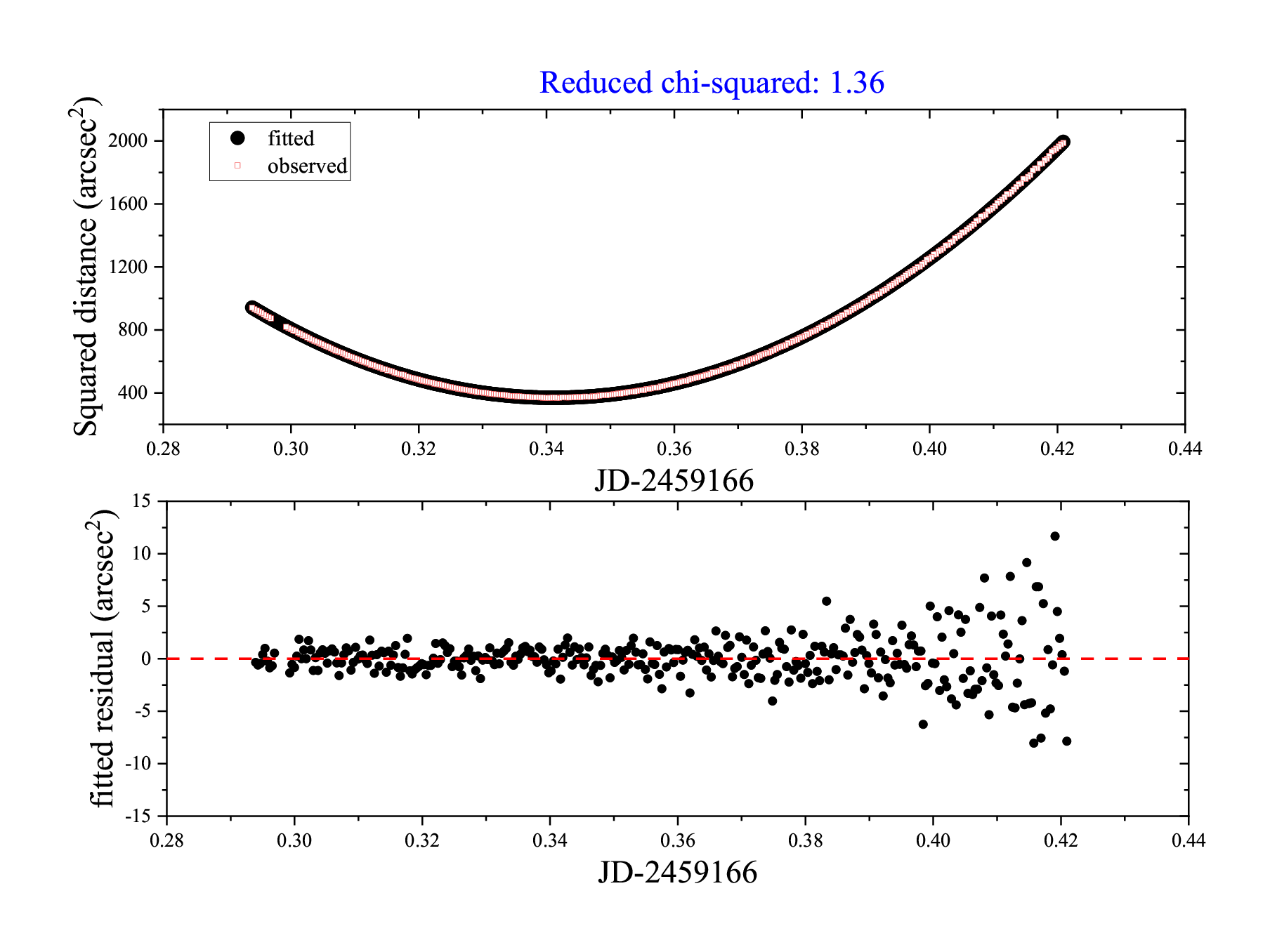}
	\caption{The figure shows the distance curve and fitted residuals of $ d^2 $ with time. The reference star ID is 190945106293176064, approached on Nov 12, 2020.}
	\label{fig:final_6064}
\end{figure}

\begin{figure}
	\centering
	\includegraphics[width=0.46\textwidth, angle=0]{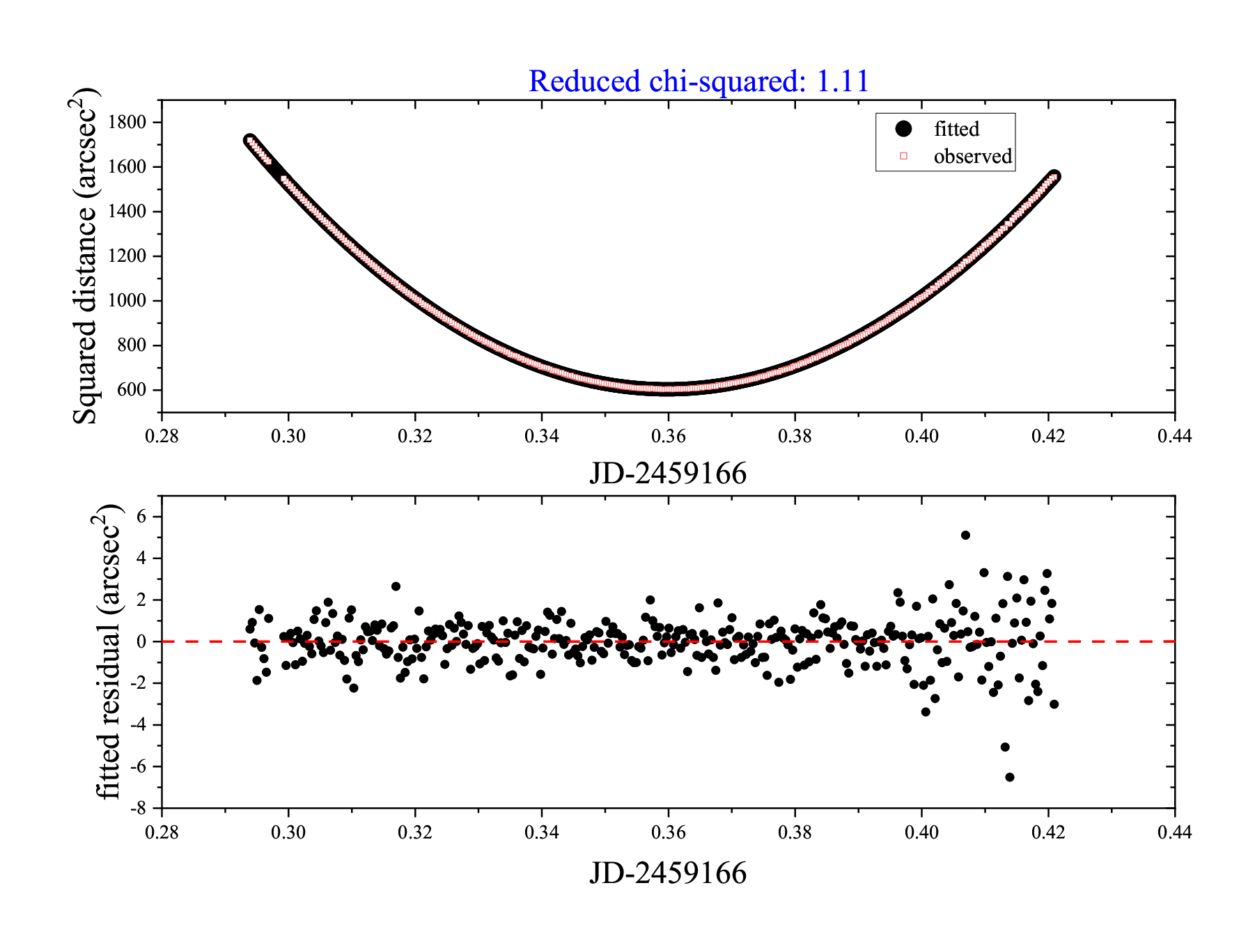}
	\caption{The figure shows the distance curve and fitted residuals of $ d^2 $ with time. The reference star ID is 190945316746612608, approached on Nov 12, 2020.}
	\label{fig:final_2608}
\end{figure}

\begin{figure}
	\centering
	\includegraphics[width=0.46\textwidth, angle=0]{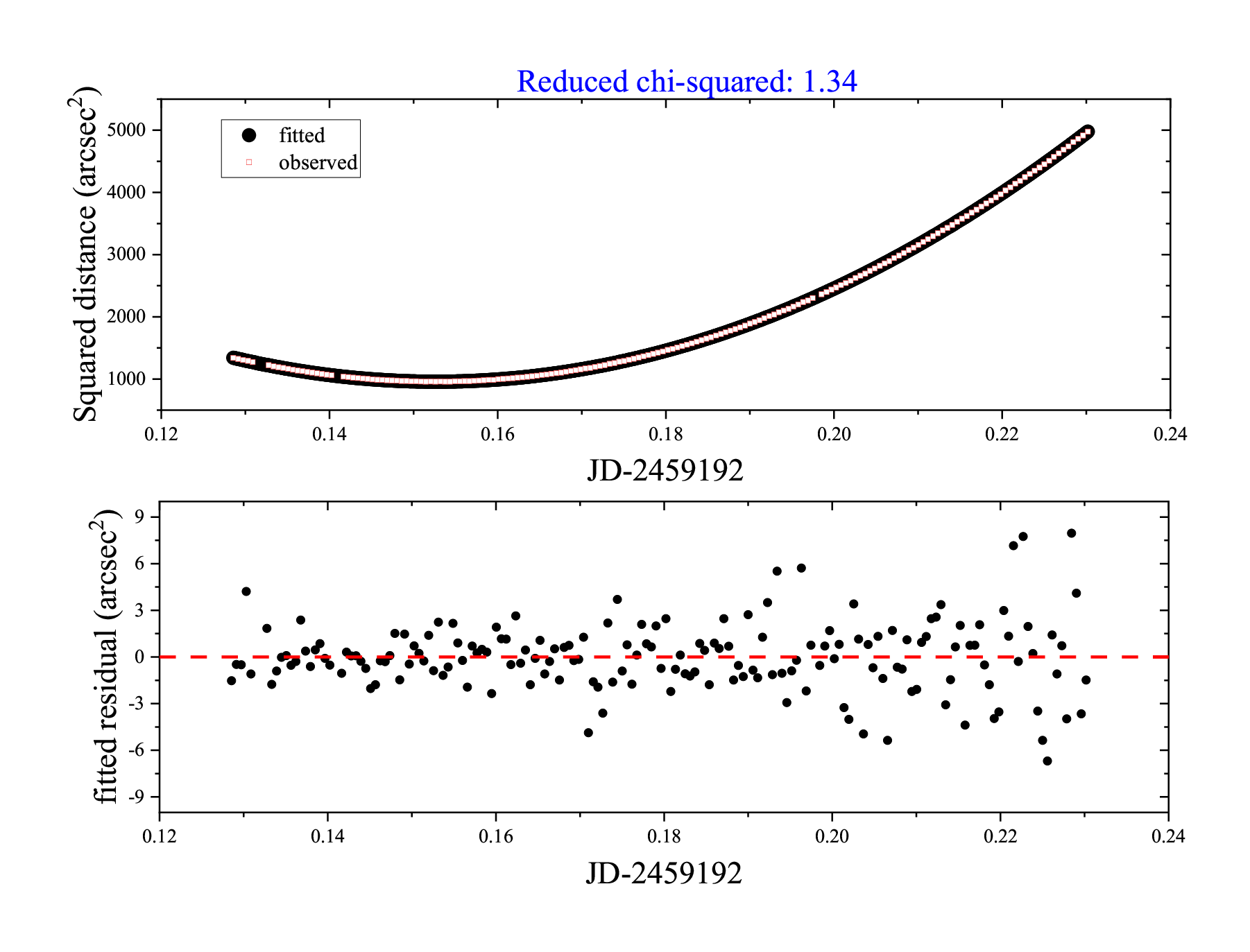}
	\caption{The figure shows the distance curve and fitted residuals of $ d^2 $ with time. The reference star ID is 188197534110556032, approached on Dec 8, 2020.}
	\label{fig:final_6032}
\end{figure}

\begin{figure}
	\centering
	\includegraphics[width=0.44\textwidth, angle=0]{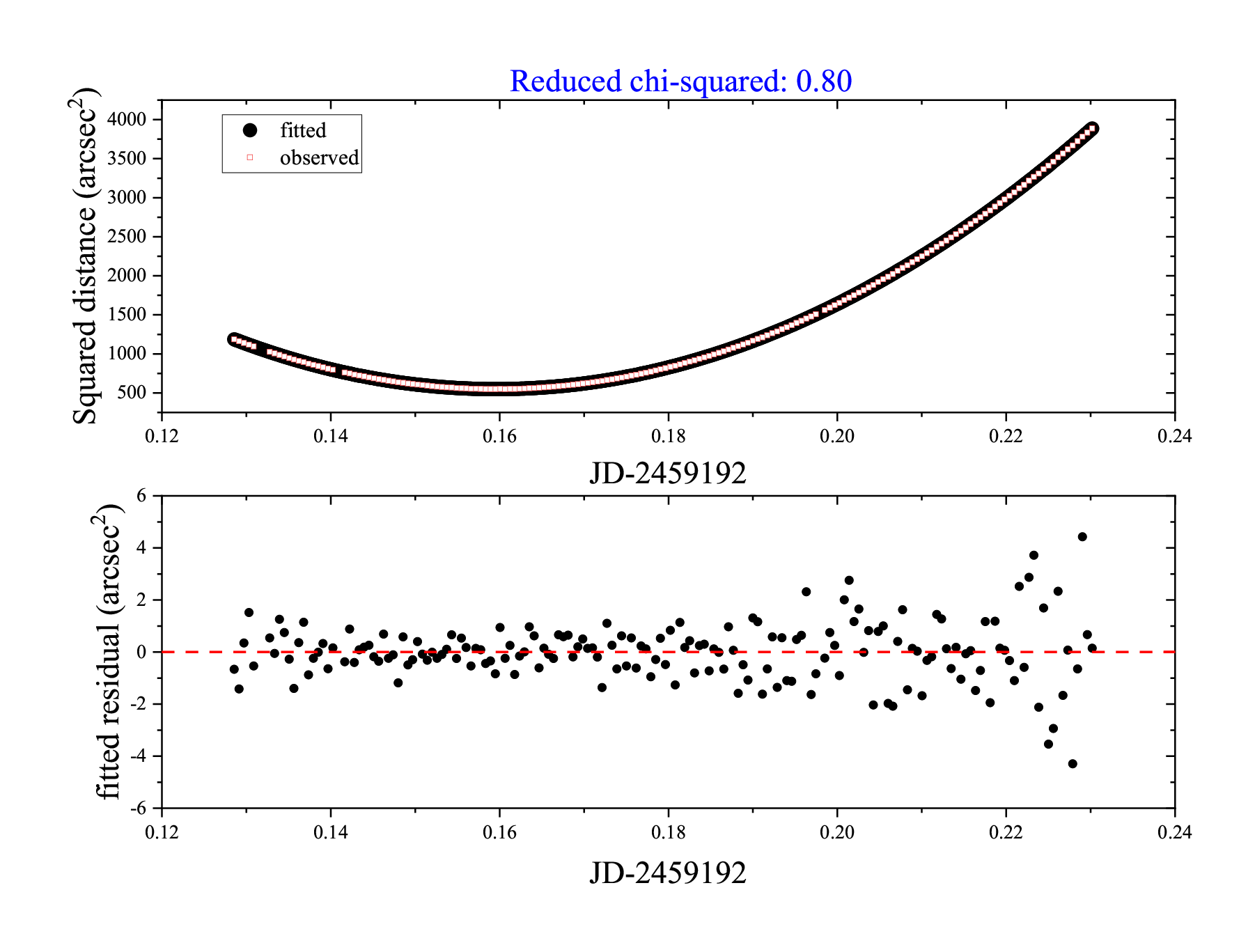}
	\caption{The figure shows the distance curve and fitted residuals of $ d^2 $ with time. The reference star ID is 188197534110557952, approached on Dec 8, 2020.}
	\label{fig:final_7952}
\end{figure}

\begin{figure}
	\centering
	\includegraphics[width=0.44\textwidth, angle=0]{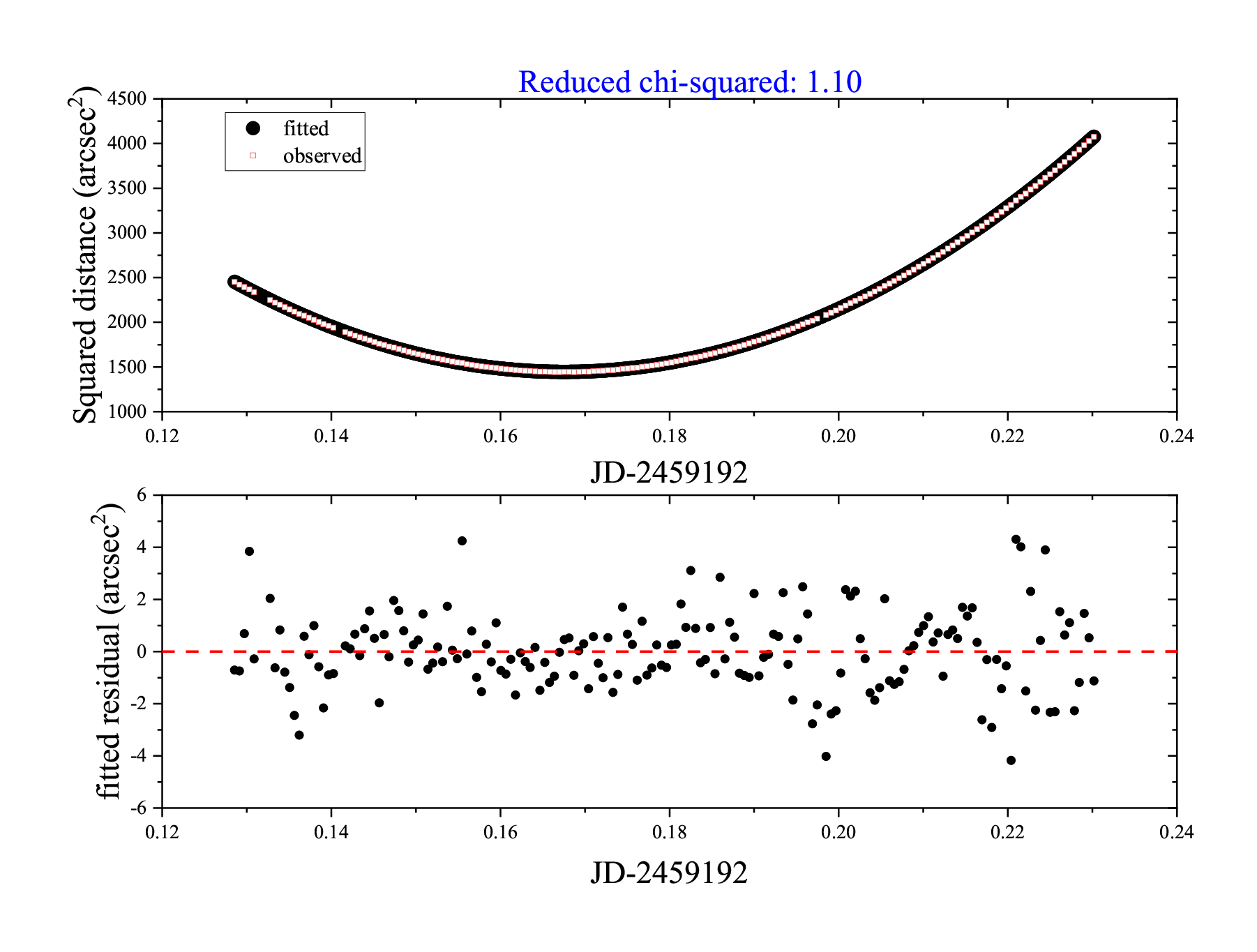}
	\caption{The figure shows the distance curve and fitted residuals of $ d^2 $ with time. The reference star ID is 188197534110557696, approached on Dec 8, 2020.}
	\label{fig:final_7696}
\end{figure}

\begin{figure}
	\centering
	\includegraphics[width=0.44\textwidth, angle=0]{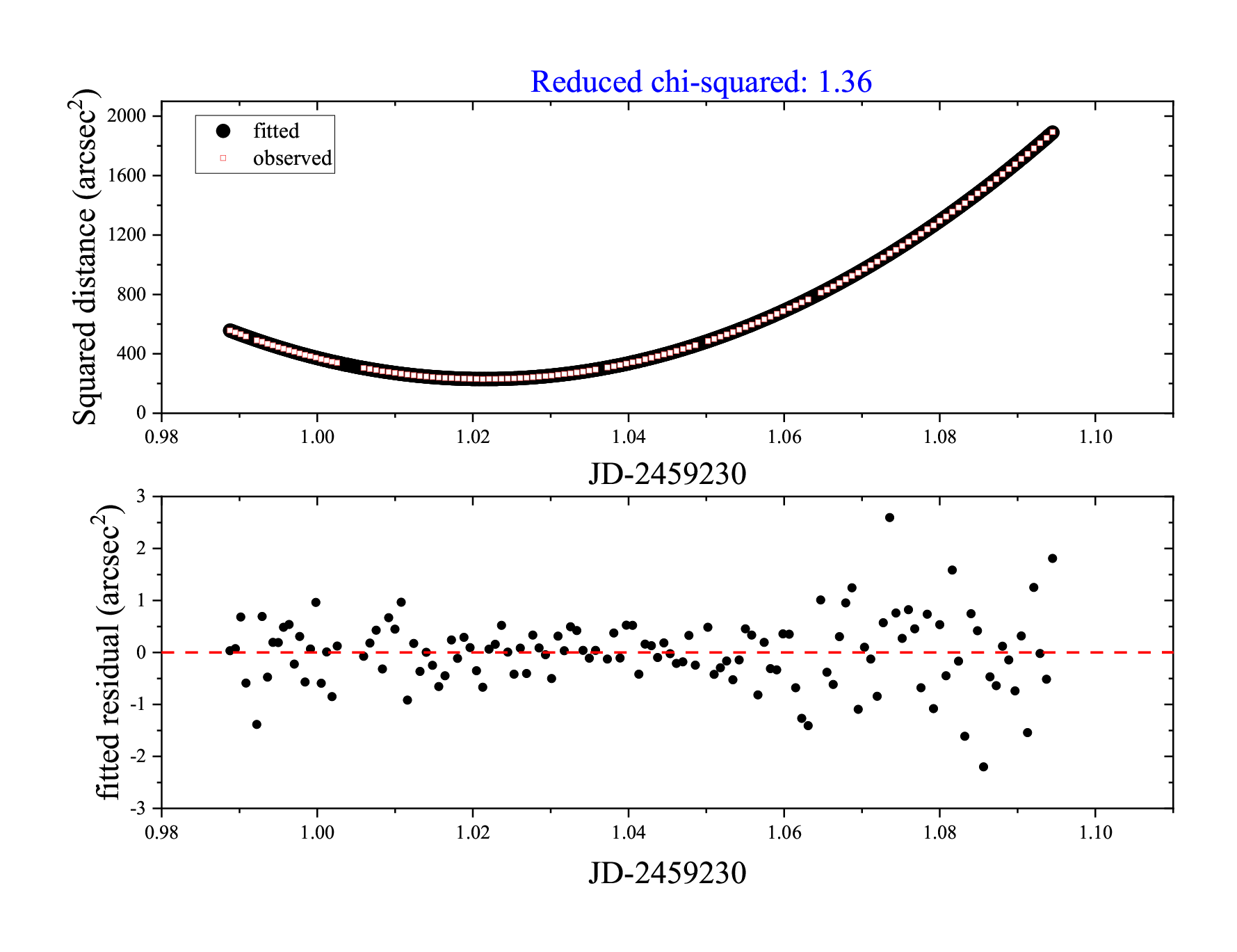}
	\caption{The figure shows the distance curve and fitted residuals of $ d^2 $ with time. The reference star ID is 173523550707799552, approached on Jan 15, 2021.}
	\label{fig:final_9552}
\end{figure}

\begin{figure}
	\centering
	\includegraphics[width=0.44\textwidth, angle=0]{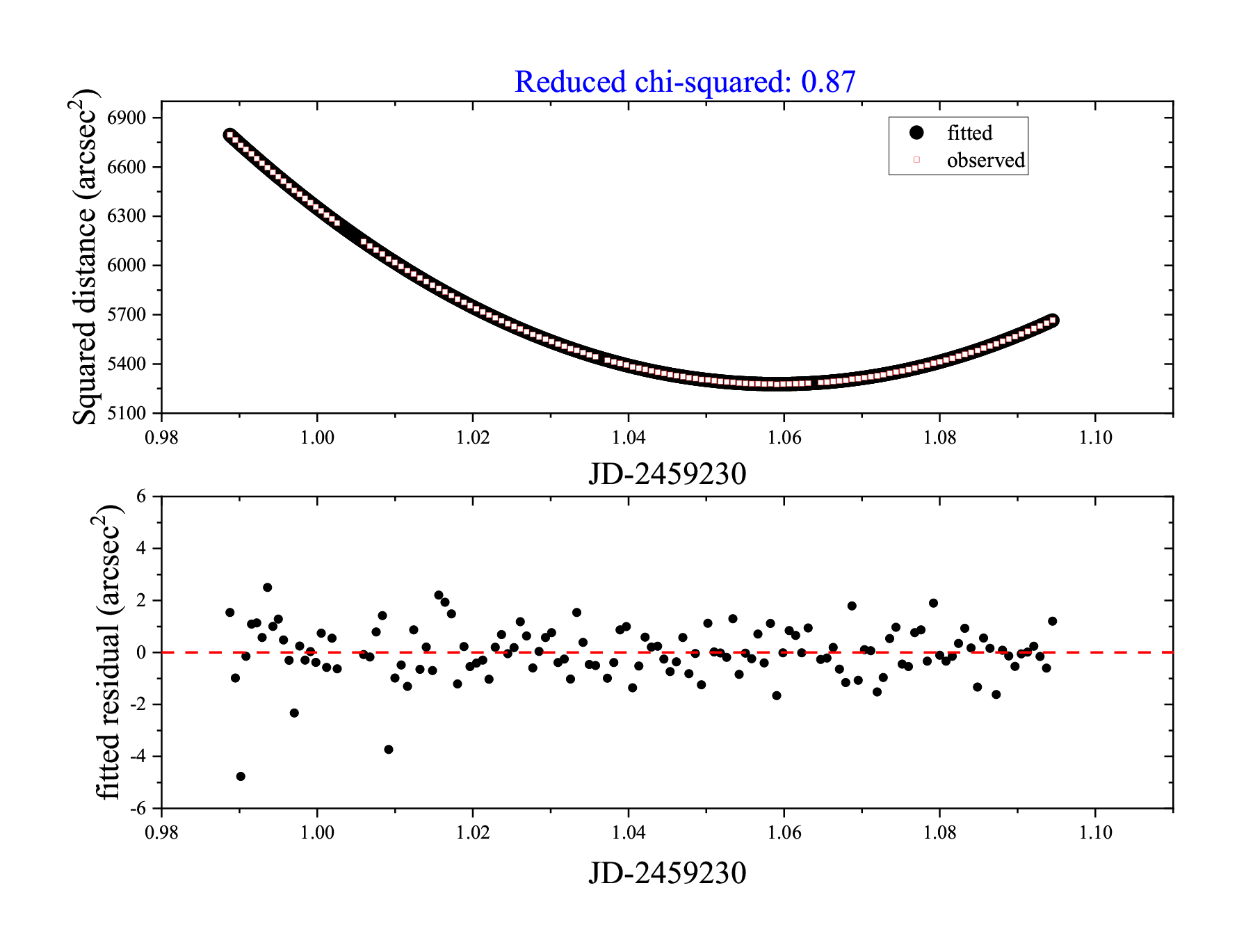}
	\caption{The figure shows the distance curve and fitted residuals of $ d^2 $ with time. The reference star ID is 173523275829562624, approached on Jan 15, 2021.}
	\label{fig:final_2624}
\end{figure}

\begin{figure}
	\centering
	\includegraphics[width=0.44\textwidth, angle=0]{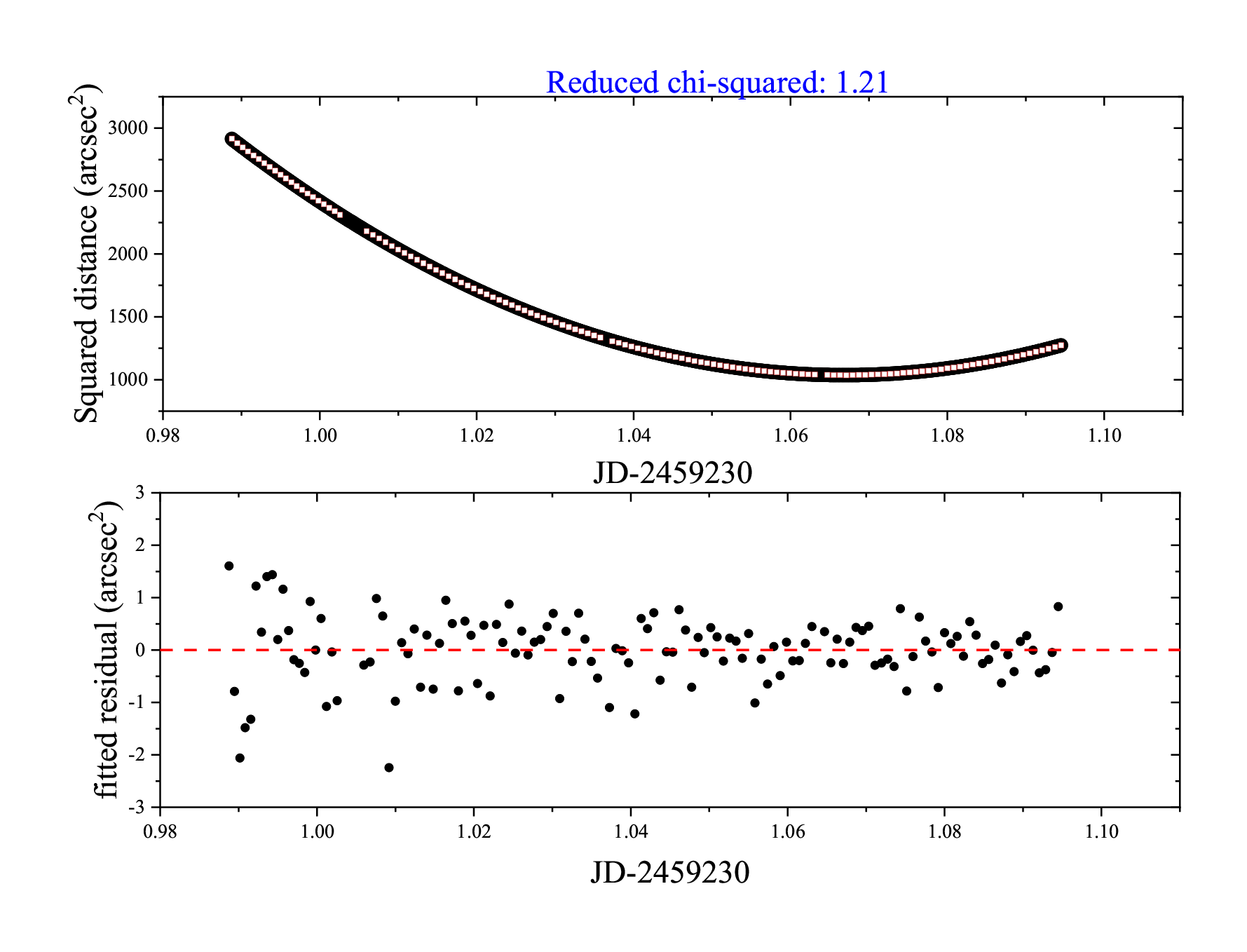}
	\caption{The figure shows the distance curve and fitted residuals of $ d^2 $ with time. The reference star ID is 173523344549362432, approached on Jan 15, 2021.}
	\label{fig:final_2432}
\end{figure}

\begin{figure}
	\centering
	\includegraphics[width=0.44\textwidth, angle=0]{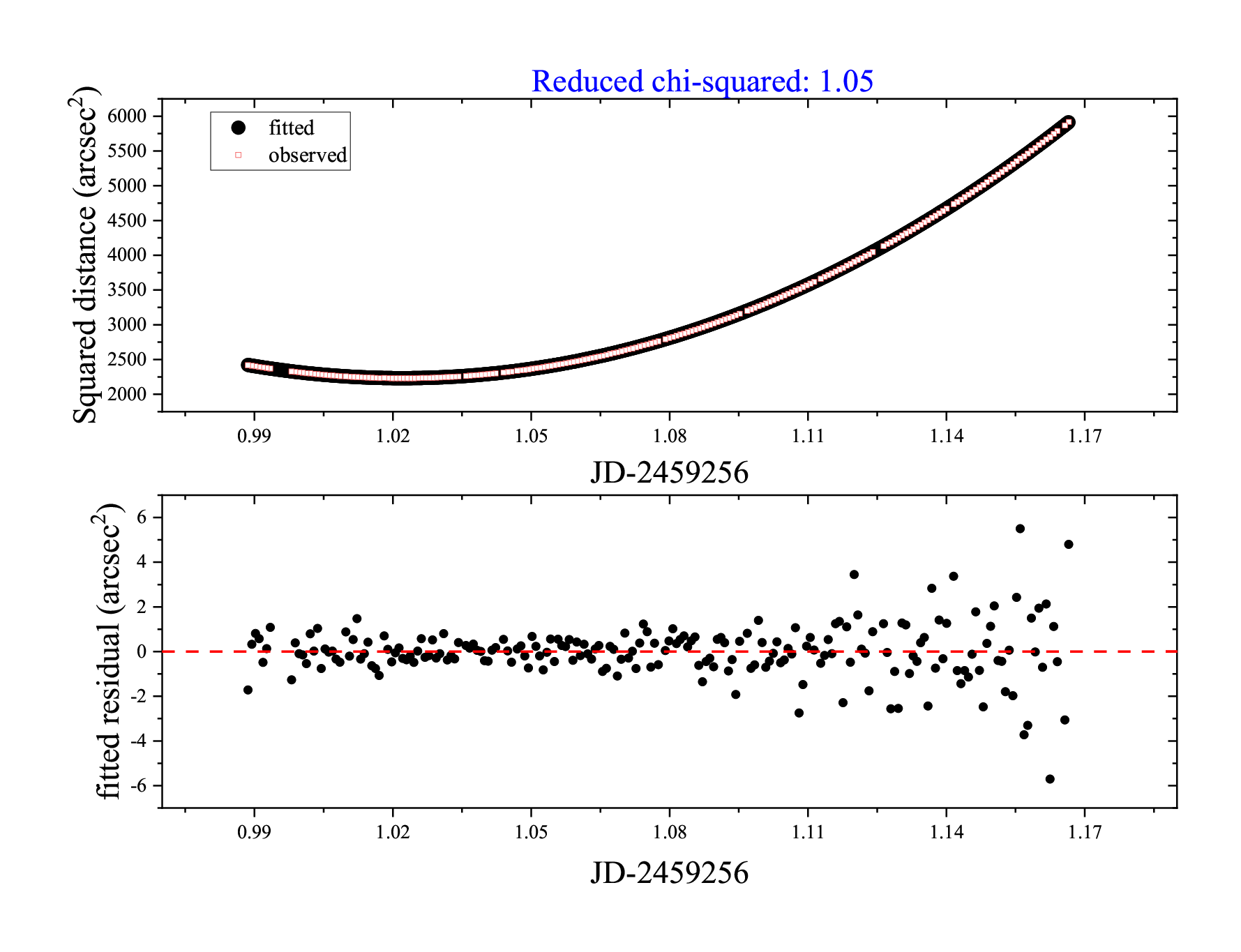}
	\caption{The figure shows the distance curve and fitted residuals of $ d^2 $ with time. The reference star ID is 159439940627025408, approached on Feb 10, 2021.}
	\label{fig:final_5408}
\end{figure}

\begin{figure}
	\centering
	\includegraphics[width=0.44\textwidth, angle=0]{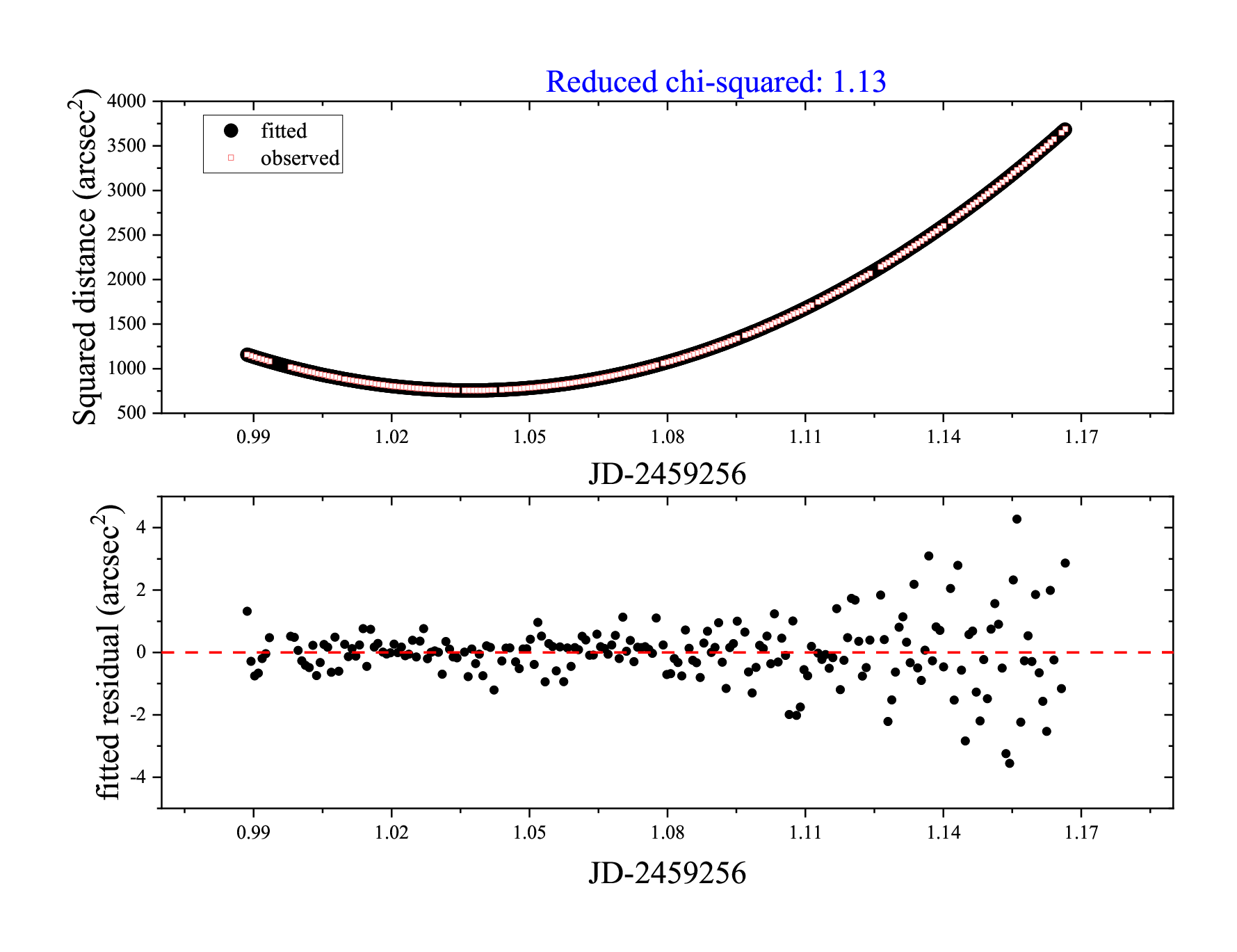}
	\caption{The figure shows the distance curve and fitted residuals of $ d^2 $ with time. The reference star ID is 159440048001339008, approached on Feb 10, 2021.}
	\label{fig:final_9008}
\end{figure}

\begin{figure}
	\centering
	\includegraphics[width=0.44\textwidth, angle=0]{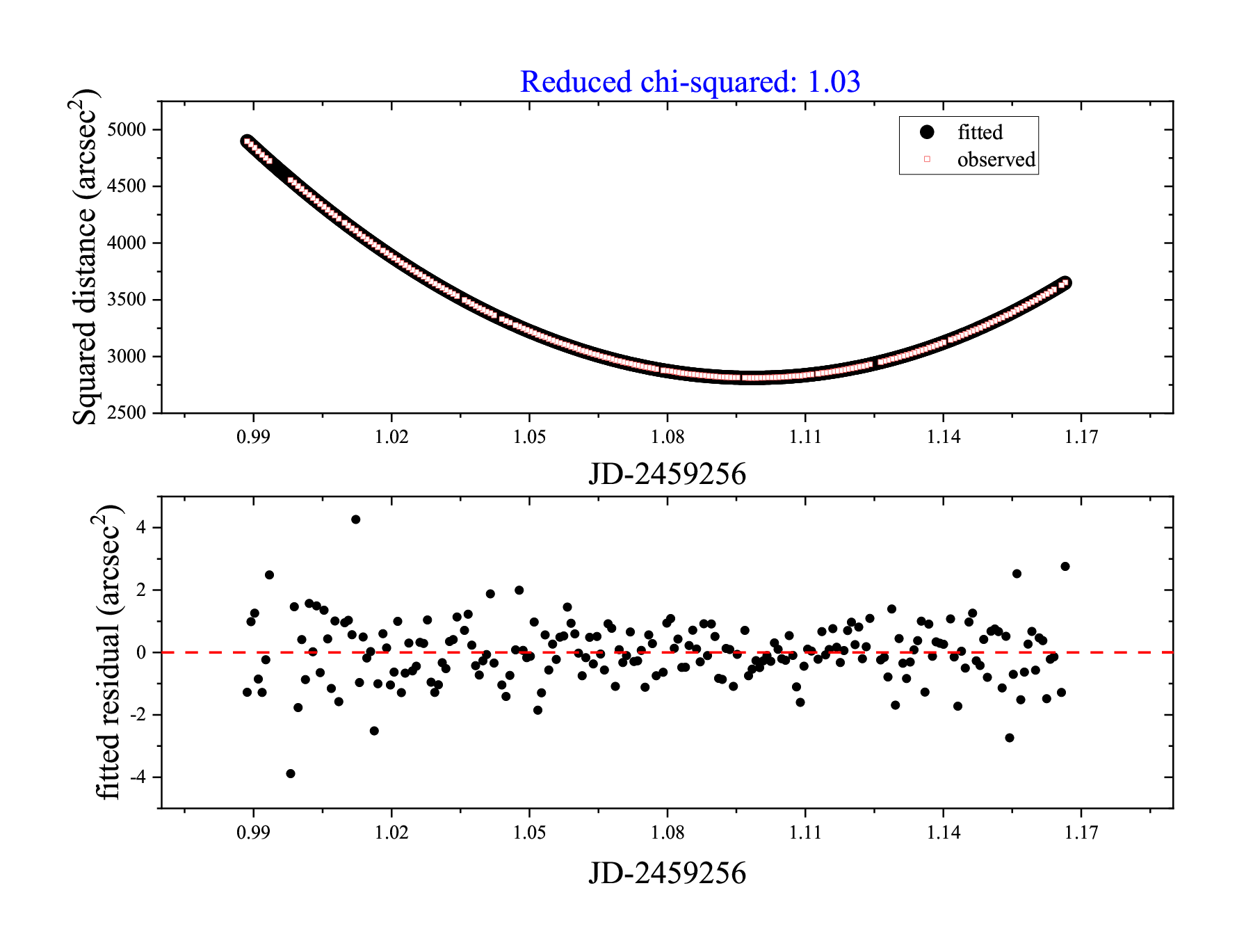}
	\caption{The figure shows the distance curve and fitted residuals of $ d^2 $ with time. The reference star ID is 159439910562389760, approached on Feb 10, 2021.}
	\label{fig:final_9760}
\end{figure}

\begin{figure}
	\centering
	\includegraphics[width=0.44\textwidth, angle=0]{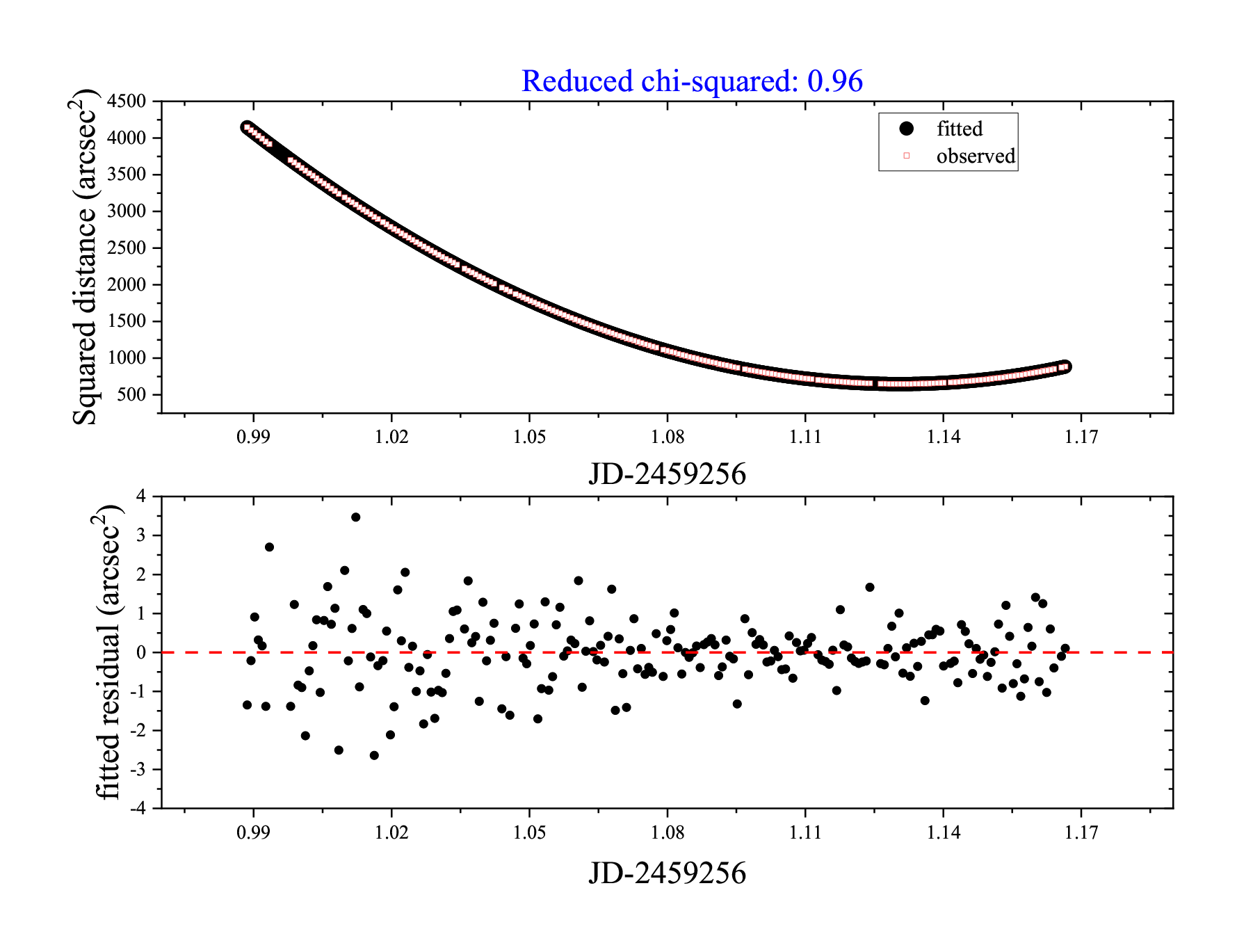}
	\caption{The figure shows the distance curve and fitted residuals of $ d^2 $ with time. The reference star ID is 159439738763696640, approached on Feb 10, 2021.}
	\label{fig:final_6640}
\end{figure}


\bsp	
\label{lastpage}
\end{document}